# Mechanics-Informed Machine Learning for Geospatial Modeling of Soil Liquefaction: Global and National Surrogate Models for Simulation and Near-Real-Time Response

Morgan D. Sanger*[1], Mertcan Geyin[2], and Brett W. Maurer[3]

**Abstract:** Using machine learning (ML), high performance computing, and a large body of geospatial information, we develop surrogate models to predict soil liquefaction across regional scales. Two sets of models – one global and one specific to New Zealand – are trained by learning to mimic geotechnical models at the sites of in-situ tests. Our geospatial approach has conceptual advantages in that predictions: (i) are anchored to mechanics, which encourages more sensible response and scaling across the domains of soil, site, and loading characteristics; (ii) are driven by ML, which allows more predictive information to be used, with greater potential for it to be exploited; (iii) are geostatistically updated by subsurface data, which anchors the predictions to known conditions; and (iv) are precomputed everywhere on earth for all conceivable earthquakes, which allows the models to be executed very easily, thus encouraging user adoption and evaluation. Test applications suggest that: (i) the proposed models outperform others to a statistically significant degree; (ii) the geostatistical updating further improves performance; and (iii) the anticipated advantages of region-specific models may largely be negated by the benefits of learning from larger global datasets. These models are best suited for regional-scale liquefaction hazard simulation and near-real-time response and are accompanied by variance products that convey where, and to what degree, the ML-predicted liquefaction response is influenced by local geotechnical data.

## Introduction

Reliable predictions of soil liquefaction are desired both prior to an earthquake for planning and mitigation, and immediately after for informing response and recovery. Such predictions could ideally be made: (i) quickly, as in near-real-time after an event; (ii) at high resolution, consistent with the scale of infrastructure assets or property parcels; and (iii) over the areal extents impacted by large earthquakes, such as those of a city or transportation network. However, because state-of-practice liquefaction models require subsurface geotechnical data such as cone penetration test (CPT) measurements, they cannot be implemented continuously across a large area. As a result, models that use "geospatial" proxy variables are often used in regional-scale studies. These variables include metrics of topography, geology, hydrology, geomorphology, ecology, groundwater, and climate that are available from existing maps, models, and remote-sensing datasets, and which correlate with traits pertinent to liquefaction (e.g., soil thickness, density, saturation, and typology). By way of these variables, geospatial models essentially infer subsurface conditions without subsurface measurements and rapidly predict liquefaction at any number of locations.

[1] Department of Civil & Environmental Engineering University of Washington, 3760 E. Stevens Way NE Seattle, WA 98195, sangermd@uw.edu, *Corresponding author
[2] Norwegian Geotechnical Institute, 1 Beacon St, Boston, MA 02108, USA, mertcan.geyin@ngi.no
[3] Department of Civil & Environmental Engineering University of Washington, 3760 E. Stevens Way NE Seattle, WA 98195, bwmaurer@uw.edu

Models of this type have existed for several decades ("HAZUS" from the National Institute of Building Science (1997) was among the first), yet they have gained new attention in recent years, driven by advances in liquefaction observation data, geospatial variables, and empirical learning techniques (e.g., machine learning, ML). For example, the seminal model of Zhu et al. (2017), later updated by Rashidian and Baise (2020), is adopted by the United States Geological Survey (USGS) as part of their post-earthquake informational products (e.g., Allstadt et al., 2022). It is difficult, of course, to predict liquefaction without site-specific subsurface data, especially across the diverse environments and geologic conditions found globally. Recent tests of geospatial and CPT-based liquefaction models demonstrate the promising potential of geospatial data, as well as clear shortcomings in current models (Geyin et al., 2020). Studies that utilize geospatial models for predicting liquefaction have since increased – in local and global contexts – and include efforts in Australia (Jena et al., 2023), New Zealand (e.g., Lin et al., 2021; Azul et al., 2024), the United States (Geyin et al., 2023; Bullock et al., 2023), Turkey (Asadi et al., 2024), the European Union (e.g., Bozzoni et al., 2021; Todorovic and Silva, 2022), and Korea (Kim, 2023), among others.

While recent literature has grown the science and adoption of geospatial models, we contend existing models share one or more significant limitations. *First*, they tend to directly predict outcomes (i.e., liquefaction manifestations, or lack thereof) without explicit consideration of, or insights into, the mechanistic causes of those outcomes. Liquefaction is best predicted by mechanics and much has been learned of these mechanics over the last 50 years. This knowledge is continually embedded in state-of-practice geotechnical models, yet geospatial models tend not to learn from, or anchor to, these mechanistic models in any way. The lack of a mechanistic backbone could be overcome with enough training data, such that a model "relearns" the governing mechanics by way of observed outcomes, but current liquefaction inventories are arguably too sparse, with data from perhaps one earthquake annually. As a result, geospatial models can depart from mechanistic principles, especially in poorly populated regions of their parameter spaces, such that a model may predict liquefaction when the shaking intensity or duration is easily judged by an expert as insufficient (e.g., Allstadt et al., 2022, Figure 4d). In some cases, ad-hoc corrections have been used to limit the misgivings of data sparsity. Allstadt et al. (2022), for example, modified the Zhu et al. (2017) model by capping input variables and by applying a magnitude scaling function, which can substantially alter the model's output (e.g., see Allstadt et al. 2022, Figure 4e). The problem posed by geospatial modeling might thus ideally be parsed into the empirical (prediction of subsurface engineering properties, conditioned on geospatial variables) and the mechanistic (prediction of liquefaction effects, conditioned on engineering properties).

*Second*, geospatial models tend not to be updated by subsurface geotechnical data. Because geospatial predictions implicitly infer subsurface traits, they could presumably be improved with direct measurements of those traits. Geotechnical data is increasingly accessible in regional and national community databases, some of which already contain hundreds of thousands of tests (e.g., New Zealand EQC, 2016; Kwak et al., 2021; Ulmer et al., 2023; Washington State DNR, n.d.). These data are likely to grow indefinitely, through continual testing and increased availability of historic data. When input to geotechnical models, geotechnical data can produce predictions of liquefaction that differ greatly from those of their geospatial counterparts. The lack of communication between geotechnical and geospatial data and models is a significant lost opportunity. Geospatial models would undoubtedly benefit from consideration of subsurface data, where available.

*Third*, existing geospatial models tend to use relatively few of the publicly available geospatial variables. Rashidian and Baise (2020), for example, use five. One variable represents demand (peak ground velocity, *PGV*) and four represent capacity (distance to water, mean annual precipitation, and the expected groundwater depth and shear-wave velocity over the upper 30 m ($V_{S30}$)). While these inputs seem to model liquefaction hazards with relative sufficiency, at least in certain locations and events, more predictive information is needed to further improve model performance and portability (i.e., stability across events, regions, and subsurface conditions). As an example, Geyin et al. (2020) observed that the inability to infer soil typology (and thus, liquefaction susceptibility) was a common cause of geospatial mispredictions.

*Fourth*, existing geospatial models tend to be trained by traditional statistical methods (e.g., logistic regression). In this regard, the potential of geospatial modeling – using many variables that weakly correlate to subsurface traits in nonlinear, interrelated ways – may not be fully realized. Logistic regression requires: (i) hypotheses of what variables matter and how; (ii) little-to-no correlation between variables; and (iii) linearity between variables and targets, which often deviate from reality where behaviors are nonlinear. Better predictions might be realized using emergent "artificial intelligence (AI)" techniques (such as ML), which could allow more predictive information to be used, with greater potential for that information to be exploited. Todorovic and Silva (2022), for example, trained an ML model to directly predict liquefaction observations using several geospatial variables, similar to Rashidian and Baise (2020), and showed evidence of improvement in unbiased tests.

*Fifth*, although AI brings opportunity to geotechnical engineering, existing AI liquefaction models are rife with problems (Maurer and Sanger, 2023). Most problematic with respect to the scientific process is that AI models are rarely provided. A large majority of publications describe the development and performance of an AI model, but do not "define" the model (i.e., do not provide code, software, or any means of use), meaning it cannot be applied or tested by anyone. Beyond this immediate concern is another specific to geospatial models: are they feasible to implement? Consider a hypothetical model that uses many high-resolution variables. Deploying such a model globally could require compilation and storage of hundreds, if not thousands, of gigabytes of geospatial data. The model itself could also be unwieldy in size.

This presents a barrier to all but large enterprises if the required data cannot easily be downloaded and stored locally. It may be argued that even geospatial models with only a few variables do not lend themselves well to adoption and testing, given that those variables must be individually located (often from broken hyperlinks) and may need to be globally computed from other raw data. If a model is not easily implemented, it will not be adopted, tested, or improved. It is thus critical to package data for users, build software interfaces, and/or develop modeling strategies that circumvent the problem of size altogether. AI models will otherwise not be used.

In this paper, we train geospatial liquefaction models (henceforth GLMs) that directly address each of these limitations using an approach that is very different from others, and which builds on concepts introduced in Geyin et al. (2022). Rather than predict liquefaction observations directly, we train geospatial "surrogate" models to mimic the predictions of geotechnical models at sites of in-situ tests. By anchoring to mechanistic models, the geospatial models benefit from the knowledge embedded therein. This encourages more sensible model response and scaling across the domains of soil, site, and earthquake loading characteristics. The predictions are made using a very large library of geospatial information, are trained using ML techniques, and are geostatistically updated in the vicinity of subsurface data, such that the geospatial models are brought into agreement with geotechnical predictions where available. Furthermore, the models are designed for ease of use. This is accomplished by effectively precomputing the expected liquefaction response at every location on earth for all potential earthquakes. This response is stored as mapped parameters that await ground-motion information from a specific earthquake (e.g., one that has just occurred, or a scenario event of interest). When convolved, these inputs rapidly produce probabilistic predictions of liquefaction impacts, giving the model near-real-time capability without requiring high-performance computing (HPC) resources nor advanced modeling capabilities. We develop surrogate ML models for several geotechnical models, such that their predictions can be ensembled, and we explore the prospects of region-specific GLMs by developing one in New Zealand. The projected benefits of our approach are further developed in *Data and Methodology*.

**Data and Methodology**

*Subsurface Data and Geotechnical Predictions as Model Targets*

Subsurface data – and geotechnical model predictions using these data – underpin our approach. These are used both to train the GLMs and to subsequently anchor their predictions to reality, such that predictions near in-situ tests are updated by (i.e., brought in closer agreement with) geotechnical models. By transferring the prediction target from liquefaction observations to subsurface data, the potential training set becomes orders-of-magnitude larger and samples Earth's terrain more broadly. This is because the sites of in-situ tests do not need to have experienced an earthquake (i.e., be liquefaction "case histories") but merely require data that can be input to a state-of-practice liquefaction model (currently CPT, $V_S$, or standard penetration test (SPT) measurements). Given the rise of community data, and research policies and infrastructures that reward data sharing (e.g., Baker et al., 2024), the disparity between the number of geotechnical tests and the number of liquefaction observations will only increase. This should allow the models developed herein to be retrained and improved more frequently, whereas geospatial models that train directly on liquefaction observations may advance less rapidly, with new data from at most a few earthquakes annually, each subjecting sample sites to just one level of seismic loading.

In this study we focus on CPTs and compile ~37,000 total tests from 48 U.S. states and 19 countries, as mapped in Fig. 1. Sources include prior international compilations (Geyin and Maurer, 2021a) and existing databases in Italy (Regione Emilia-Romagna, 2024), New Zealand (New Zealand EQC, 2016), and the United States (USGS, 2019). Considerable data were also newly compiled for this project from several thousand analog sources – focusing on North America – and are digitally available from Sanger et al. (2024a) and Rasanen et al. (2024). Although this collection of data is in many ways unprecedented, some regions of interest are still poorly represented and additional data is needed, as always, while other regions are data rich (e.g., Italy, New Zealand, United States), evoking questions of model bias. To address these issues, this study includes: (i) parameter distributions comparing the training set with global conditions; (ii) several types of unbiased model tests; and (iii) maps depicting the degree to which model predictions are influenced by geotechnical data. Additionally, the models are constrained to the training domain of select, influential parameters, meaning the models generally do not make predictions for conditions unencountered in training. These and other limitations and uncertainties are further discussed later.

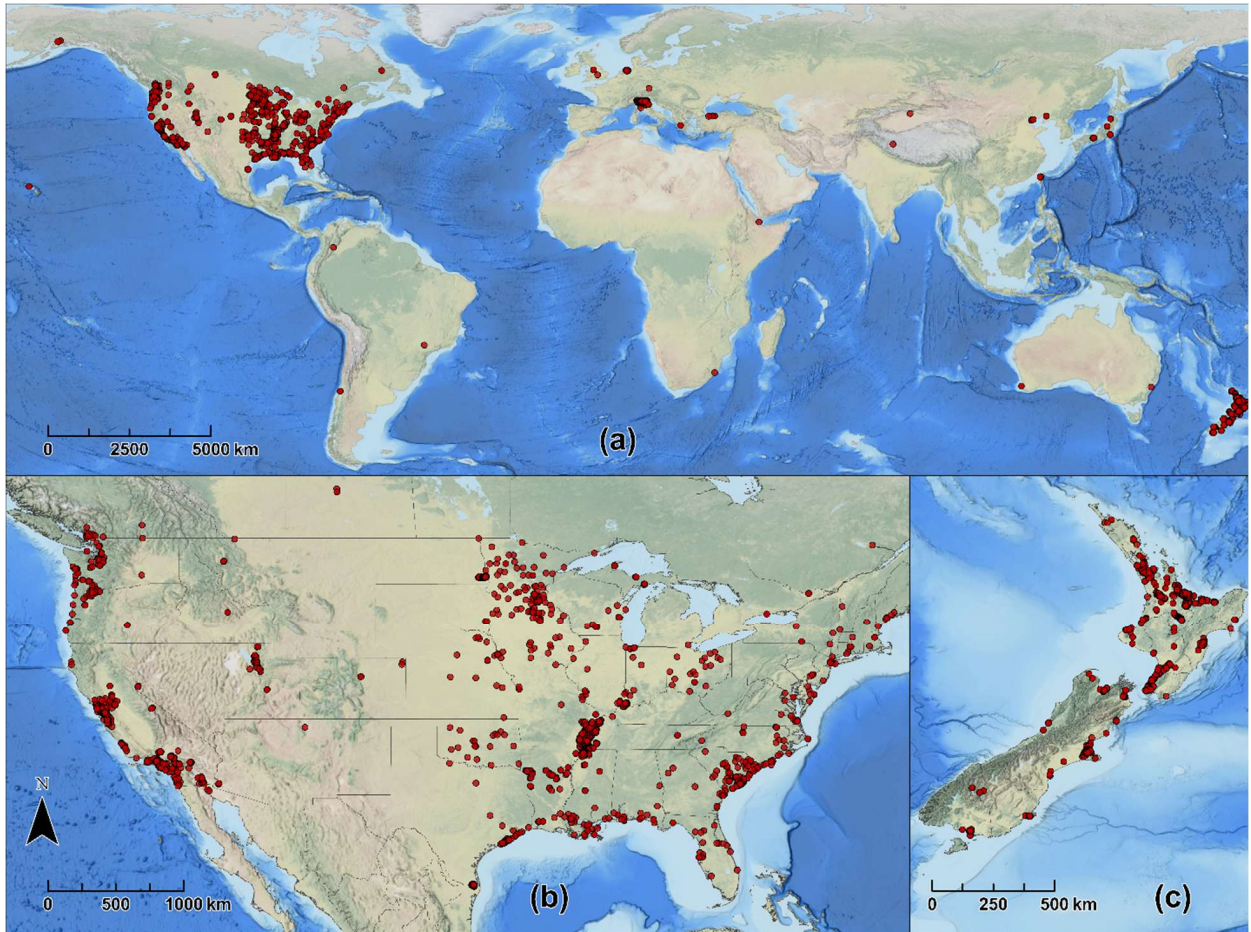

**Fig. 1.** Spatial distribution of CPT training and test data (a) globally, with a closer look at (b) the conterminous United States and (c) New Zealand. (Sources: GEBCO and NOAA NCEI.)

Each CPT was processed and standardized to interpolate CPT measurements on a constant depth interval, align tip and sleeve measurements via cross-covariance, correct noise by infilling invalid measurements with linear interpolation, and infill measurements between the ground surface and predrill depth with data from just below the predrill depth. Then, each profile was subjected to a loading array defined by peak ground accelerations (*PGA*s) of 0.05 g to 2.0 g and rupture magnitudes of 4.5 to 9.0. For each loading, the factor-of-safety against liquefaction triggering was computed as a function of depth using the Idriss and Boulanger (2008) model, which has been shown, to a statistically significant degree, to perform at least as well as all other models common in practice (Geyin et al., 2020). Soil fines content was estimated from the CPT soil-behavior-type-index, $I_c$ (Robertson, 2009), via the Boulanger and Idriss (2016) model, except in New Zealand where the regional model of Maurer et al. (2019) was used. Soils with $I_c$ >2.5 were assumed not susceptible to liquefaction per Maurer et al. (2019), where $I_c$ = 2.5 was identified as the median (i.e., 50th percentile) liquefaction susceptibility threshold per the Boulanger and Idriss (2006) criterion. Corrections for CPT volume-averaging effects (e.g., Boulanger and DeJong, 2018) were not applied based on the findings of Geyin and Maurer (2021b) and Yost et al. (2021). To predict manifestations, or consequences, of liquefaction at the ground surface, the results from triggering analysis were input to three models: the liquefaction potential index (*LPI*) (Iwasaki et al., 1978); a modified *LPI*, termed $LPI_{ISH}$ (Maurer et al., 2015a); and the liquefaction severity number (*LSN*) (van Ballegooy et al., 2014). These models, which each output an index (often called a "vulnerability index"), are used in land-use planning, hazard mapping, and engineering site-assessment to predict a soil profile's cumulative liquefaction response, or damage potential, at the ground surface. Fragility functions conditioned on *LPI*, $LPI_{ISH}$, and *LSN* have been trained using case-history data to predict the probabilities of certain outcomes, including ground failure (i.e., deformation and ejecta) (Geyin and Maurer, 2020), pipeline rupture (Toprak et al., 2019), and foundation damage (Maurer et al., 2025). Because *LPI*, $LPI_{ISH}$, and *LSN* are well known in the literature and available in engineering software (e.g., *CLIQ* by GeoLogismiki; *Design Studio* by Infinity Studio), their formulae are omitted here but provided in the *Supplemental Materials*.

Shown in Fig. 2 are the resulting *LPI* values at four of the ~37,000 CPT sites, plotted as a function of magnitude-scaled *PGA* ($PGA_M$). The relationship between *LPI* and $PGA_M$ is a unique signature of each site, with no two sites having identical responses. If it were possible to obtain this signature remotely (i.e., without in-situ data), then the expected liquefaction response across all levels of loading would, in effect, be predicted. To that end, we fit a simple but flexible functional form to these data:

$$MI(PGA_M) = \begin{cases} 0, for\ PGA_M < 0.1g \\ A * (\tan^{-1}(B * (PGA_M - \frac{A/100}{B})^2)), for\ PGA_M \geq 0.1g \end{cases} \quad \text{(Eq. 1)}$$

where *MI* is the manifestation index (i.e., *LPI*, $LPI_{ISH}$, or *LSN*), $PGA_M$ is as previously defined, *A* and *B* are independent fitting parameters that will subsequently be predicted by ML, and $\tan^{-1}$ is expressed in radians. A threshold $PGA_M$ of 0.1 g was adopted in the piecewise function to maximize its fit of the data across the $PGA_M$ domain; computed *MI* values were very rarely non-zero below this threshold, which is consistent with global observations that manifestations have near-zero probability at *PGA* < 0.1 g (e.g., de Magistris, 2013).

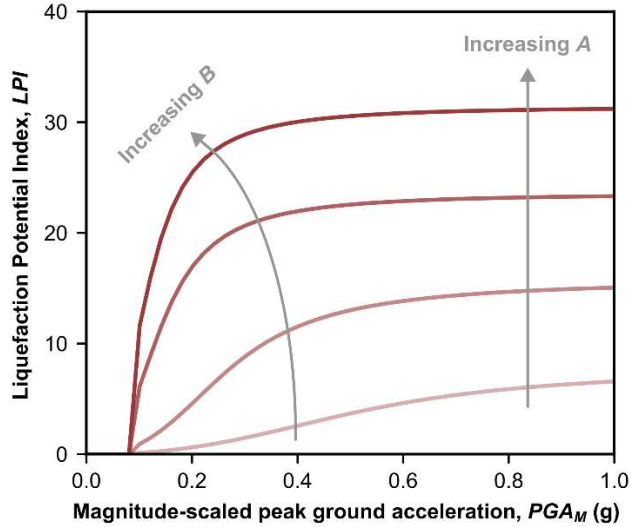

**Fig. 2.** Example *LPI* versus $PGA_M$ curves for four sites.

Eq. (1) is expressly formulated so that *A* and *B* may be stored in 16-bit format, which substantially compresses the size of eventual model products. Although *A* and *B* lack exact physical meaning, *A* generally describes the *MI* attained at relatively high $PGA_M$ and *B* generally captures the sensitivity of the liquefaction response to ground shaking (i.e., how fast *MI* reaches its maximum value). These fitting parameters exhibit very weak to weak statistical correlation and are therefore modeled independently. The fitting of Eq. (1) to the *MI* data results in a fitting error, or uncertainty, that is normally distributed, unbiased, and very small for most sites. We view this uncertainty as negligible compared to others, and when considering what Eq. (1) conceptually permits: a geospatial prediction of liquefaction response that is mechanics-informed and updatable using geotechnical data, for all seismic loading. And, because *A* and *B* are event-independent, they can be globally predicted in advance of their use, at which time *LPI*, $LPI_{ISH}$, and *LSN* are computable at low computational cost. An added advantage of this modularity is that users can pair these predictions with fragility functions of their choice, making it feasible to tailor products for different needs and utilize the latest models, since fragility functions are frequently proposed or updated using new case-history data.

*Geospatial Predictors as Model Features*

Whereas liquefaction is best modeled considering mechanics, the relationship between geospatial variables and subsurface traits is empirical, involving many interrelated correlations. Domain expertise may guide the selection of variables, but there is no expectation these variables will relate to soil traits in a mechanistic way. In the current effort, two sets of models are developed to predict *LPI*, *LPI$_{ISH}$*, and *LSN* via parameters *A* and *B*: (i) a global model, meaning predictors must be globally available; and (ii) a model specific to New Zealand, which will be used to judge the potential for region-specific models to perform better. Although global models can train on more data, region-specific models have two attractions: (i) conditions are likely to be more consistent (e.g., geology, geomorphology, and climate); and (ii) better predictor variables may be available, with higher resolution or more regional specificity, as compared to those with global coverage. For these reasons, the relationships between variables and targets could have less variance in a regional setting. New Zealand was chosen for this pilot because it has a large amount of geotechnical data and several region-specific predictors (e.g., national models of groundwater, $V_{S30}$, geology, and soils).

A total of 37 variables were ultimately chosen for the global models through an iterative process that considered correlation structures, measurements of variable importance and model performance, overfitting behavior, and the authors' judgement, both in selecting provisional variables and when inspecting final products. The name, spatial resolution, and source of each variable is in Table 1. Most variables are available at multiple spatial resolutions, but only one was ultimately adopted through the process above. Many other provisional variables were omitted entirely (e.g., mean annual precipitation). Table S1 in the *Supplemental Materials* provides additional information, including definitions, descriptions, and hyperlinked sources. However, the end-user is reminded that these variables are not required to execute the models (in contrast to other geospatial models), since model predictions will be stored as mapped parameters *A* and *B*.

**Table 1.** Summary of predictor variable information.

| Variable | Units | Resolution | Source |
| --- | --- | --- | --- |
| Bulk density | kg/m3 | 250 m | Hengl (2018a) |
| Clay fraction | kg/kg | 250 m | Hengl (2018b) |
| Compound topographic index (CTI) | -- | ~90 m (3 arc-sec) | Amatulli et al. (2020) |
| Convergence | -- | ~90 m (3 arc-sec) | Amatulli et al. (2020) |
| Depth to bedrock | cm | 250 m | Shangguan et al. (2017) |
| Depth to groundwater | m | ~100 m | Fan et al. (2013) |
| Distance to coast | km | ~1100 m (0.01 deg) | NASA (2020) |
| River distances (Flow orders 1-8) | m | 250 m | Lehner and Grill (2013) |
| Elevation standard deviation | m | ~90 m (3 arc-sec) | Amatulli et al. (2020) |
| Geomorphon | -- | ~90 m (3 arc-sec) | Amatulli et al. (2020) |
| Height above nearest drainage (HAND) | m | 1000 m | Nobre et al. (2011) |
| Landform entropy | -- | 1000 m | Amatulli et al. (2018) |
| Landform Shannon index | -- | 1000 m | Amatulli et al. (2018) |
| Landform uniformity | -- | 1000 m | Amatulli et al. (2018) |
| Major | -- | 1000 m | Amatulli et al. (2018) |
| Maximum multiscale deviation (MMD) | -- | ~90 m (3 arc-sec) | Amatulli et al. (2020) |
| Maximum multiscale roughness (MMR) | deg | ~90 m (3 arc-sec) | Amatulli et al. (2020) |
| Profile curvature | rad/m | ~90 m (3 arc-sec) | Amatulli et al. (2020) |
| Roughness | m | ~90 m (3 arc-sec) | Amatulli et al. (2020) |
| Sand fraction | g/kg | 250 m | Hengl (2018c) |
| Scale of the MMD | -- | ~90 m (3 arc-sec) | Amatulli et al. (2020) |
| Scale of the MMR | deg | ~90 m (3 arc-sec) | Amatulli et al. (2020) |
| Silt fraction | kg/kg | 250 m | Hengl (2018d) |
| Soil class | -- | 250 m | Hengl and Nauman (2018) |
| Tangential curvature | rad/m | ~90 m (3 arc-sec) | Amatulli et al. (2020) |
| Terrain ruggedness index (TRI) | m | ~90 m (3 arc-sec) | Amatulli et al. (2020) |
| Topographic position index (TPI) | m | ~90 m (3 arc-sec) | Amatulli et al. (2020) |
| Topographic slope | % | ~90 m (3 arc-sec) | Amatulli et al. (2020) |
| Vector ruggedness measure (VRM) | m | ~90 m (3 arc-sec) | Amatulli et al. (2020) |
| Vs30 | m/s | 100 m | Heath et al. (2020) |
| Water content | % | 250 m | Hengl and Gupta (2019) |

The variables in Table 1, which were sampled at ~37,000 CPT sites, include measurements derived from elevation models and hydrologic datasets (e.g., height above nearest drainage, compound topographic index, distance to rivers of different flow-orders), and predictions made by other models (e.g., soil class, water content, and clay fraction). The goal of these variables is to correlate to soil thickness, saturation, density, and typology. To evaluate the potential for model bias and the need for bias mitigation (e.g., data resampling, variable transformations), the variables in Table 1 were also sampled at all locations on earth and the resulting distributions were compared to those from CPT sites, as shown in Figs. S1-S37. In cases of missing geospatial data at a CPT location, which occasionally occurred near water bodies, the geospatial predictor was imputed with the nearest value. It is apparent from these comparisons that deep groundwater conditions are underrepresented at CPT locations. Each CPT was therefore duplicated and randomly assigned a new groundwater depth of up to 50 m. These synthetic data were included in training so that the ML model better understands the expected liquefaction response (as predicted by geotechnical models) across a broader spectrum of groundwater conditions. Importantly, these cases – which are relatively easy to predict once the significance of deep groundwater is learned – are never included in statistics of model performance. In other words, the training set includes synthetic data, but the training and test performance metrics will not. For the New Zealand model, 43 variables were ultimately adopted and sampled at the locations of 16,475 CPTs in New Zealand. Of these, unique variables not used in the global model are summarized in Table 2 and complete variable information is provided in Table S2. Aside from differing CPT datasets and predictor variables, the methodologies applied globally and in New Zealand are the same.

**Table 2.** Predictor variables for New Zealand that differ from the global model.

| Variable | Units | Resolution | Source |
|---|---|---|---|
| Depth to groundwater | m | ~200 m | Westerhoff et al. (2018) |
| Distance to coast | km | ~1100 m | NASA (2020) |
| River distances (Strahler orders 1 to 5) | m | ~100 m | LINZ (2020) |
| Geologic unit, Deposit Type, Age | -- | 100 m | Heron (2018) |
| Pfafstetter level basin characterization | -- | ~100 m | Lehner and Grill (2013) |
| Profile curvature | rad/m | 1000 m | Amatulli et al. (2018) |
| Roughness | m | 1000 m | Amatulli et al. (2018) |
| Soil depth | -- | ~200 m | McNeill et al. (2018) |
| Soil drainage | -- | ~200 m | McNeill et al. (2018) |
| Soil order | -- | ~200 m | McNeill et al. (2018) |
| Tangential curvature | rad/m | 1000 m | Amatulli et al. (2018) |
| TRI | m | 1000 m | Amatulli et al. (2018) |
| TPI | m | 1000 m | Amatulli et al. (2018) |
| Topographic slope | % | 1000 m | Amatulli et al. (2018) |
| VRM | m | 1000 m | Amatulli et al. (2018) |
| Vs30 | m/s | 100 m | Foster et al. (2019) |

*Model Training*

AI/ML techniques allow for more predictive information to be used and increase the potential for that information to be exploited. Simultaneously, a large majority of existing AI/ML liquefaction models have serious flaws, as documented by Maurer and Sanger (2023) who reviewed 75 such models. Among other failings, many publications: (i) did not test against any existing model; (ii) departed from best practices in model development (e.g., cross validation, unbiased test sets, tests of statistical significance); and (iii) did not provide a usable model to readers. Consequently, it is often unclear how well these models perform, why they should be adopted, and how they could even be used. We are thus keenly aware of the pitfalls with AI/ML tools and address them in our methodology.

Having compiled predictor variables at CPT sites where parameters $A$ and $B$ were obtained, the data were parsed into training (90%) and test (10%) sets. Several types of ML algorithms were used to train provisional models, including different neural networks (e.g., Islam et al., 2019), and decision tree ensembles formed by bagging (e.g., Breiman, 1996), boosting (e.g., Chen and Guestrin, 2016), or random forests (Breiman, 2001). Through this iterative process, during which the training, cross validation, and test-set performances were judged for performance and overfitting behavior, bagged decision-tree ensembles were ultimately chosen, both for the global and New Zealand models (Breiman, 1996; Breiman, 1999). Decision trees map a specific combination of inputs to an expected output by way of recursive decision forks. Because a single tree is typically not especially accurate and is prone to overfitting, trees are usually ensembled. In bagging, which is also known as bootstrap aggregating, many variants of the training set are sampled, and each is used to train a model. The outputs from the various trees are then ensembled, or averaged, to form a prediction. Owing to this resampling and averaging approach, bagging tends to reduce variance and avoid overfitting, compared to other ensembling methods. An additional advantage of decision-tree models is that they are relatively interpretable versus more convoluted model architectures.

Many model iterations were created using different loss functions, k-fold cross validation partitions, and predictor variables. The model hyperparameters were individually optimized for each of the six targets (i.e., *A* and *B* for *LPI*, *LPI*$_{ISH}$, and *LSN*) using a parallelized grid search algorithm to optimize the ten-fold-cross validation mean-square-error (MSE) (model hyperparameters are further detailed in Table S3). The adoption of the MSE loss function gives some preference to reducing major mispredictions at the possible expense of more minor mispredictions. Given the clustering of data in some locales (e.g., the Christchurch, New Zealand, metropolitan area), a heuristic weighting scheme based on spatial point density was applied. This downweighed the influence of data in Christchurch by ~50%. Although such weighting diminishes performance on the training and test sets, it was desirable in pursuit of more generalizable models. The performance, implementation, and geostatistical updating of these models is next discussed.

**Results and Discussion**

*Model Performance, Application, and Geostatistical Updating*

Using these data and methods, 12 distinct ML models were developed to predict the two fitting parameters (i.e., *A* and *B*), for each of the three geotechnical models (i.e., *LPI*, *LPI*$_{ISH}$, and *LSN*) and two study areas (i.e., global and New Zealand model datasets). A representative example of performance is shown in Fig. 3 for the global *LPI* model's *A* and *B* using its test set. Analogous figures for all models, both in training and testing, appear in Figs. S38-S49. Prediction residuals (defined throughout this paper as predicted – observed) are generally unbiased and normally distributed (Fig. 3, Table S4), and *A* is consistently predicted better than *B*. This might mean that the maximum attainable *MI* at relatively high *PGA*$_M$ (which relates more to *A*) is easier to predict than the sensitivity of the liquefaction response to *PGA*$_M$ (which relates more to *B*). Another explanation is that *A* does more to define the overall shape of the *MI*-*PGA*$_M$ curves (Fig. 2) and is thus a stronger site signature than *B*, which adds nuance to the curves but is less important overall. It is worth asking whether the models for *B* are needed, given their large variances, or if it would be acceptable to instead make *B* constant for all predictions. This may be answered using the Nash-Sutcliffe coefficient, *E* (Nash and Sutcliffe, 1970). *E* = 1.0 indicates a perfect model whereas *E* < 0 indicates that adopting a mean value for *B* would be better than using a model to predict it. In this case, *E* values for all global *B* models are positive and average 0.36. Thus, while opting not to use the *B* models might not diminish predictions of *MI* substantially, it is still better to predict *B*. Because *A* and *B* lack exact physical meaning, and because they do not have equal influence in Eq. (1), it is more informative to assess performance by predicting the final targets (e.g., *LPI*).

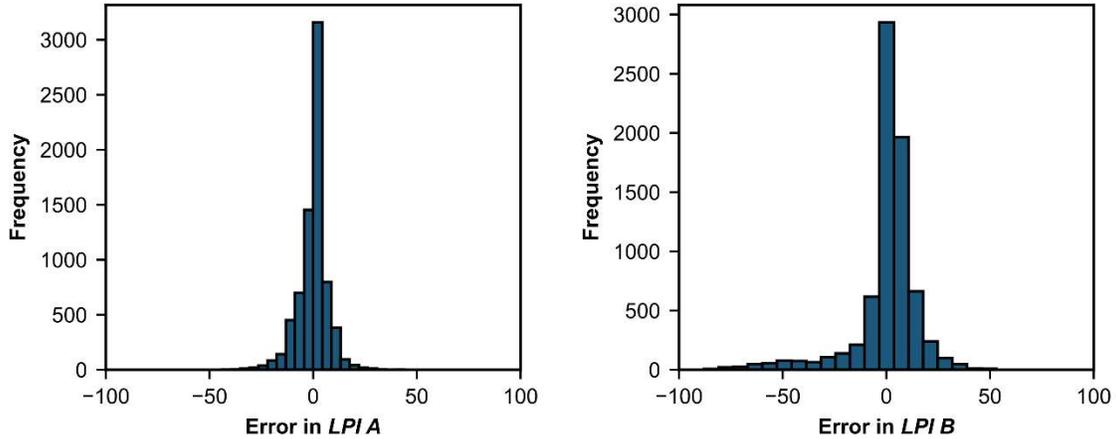

**Fig. 3.** Predicted vs. observed *LPI A* and *LPI B* for the global model test set.

In this regard, Fig. 4 illustrates *LPI* residuals as a function of $PGA_M$ for the global model test set. Across the domain $0 < PGA_M < 1$ g, these residuals have a median absolute error (MAE) of 4.5 and a median standard deviation (MSD) of 11, meaning that 68% of *LPI* prediction errors are less than ±11 and 95% are less than ±22. While readers familiar with *LPI* could initially judge these errors as being nontrivial, it is important to note that: (i) large errors are predominantly associated with large *LPI* targets; and (ii) errors in *LPI* become less consequential as *LPI* increases. According to the fragility functions of Geyin and Maurer (2020), for example, which predict the probability of ground failure (*PGF*) conditioned on *LPI*, an error of 20 is less consequential at *LPI* = 30 than an error of 2 at *LPI* = 3. This is because the expected likelihoods and severities of liquefaction manifestations become relatively constant at large *LPI*. The same is true of $LPI_{ISH}$ and *LSN*, and for this reason, errors are best interpreted after transformation to consequence predictions (i.e., by predicting outcomes conditioned on these indices). In this context, the MAE and MSD of 4.5 and 11 equate to errors in *PGF* of 8% and 22%, respectively, per Geyin and Maurer (2020), meaning that 68% of *PGF* prediction errors are within ±22%. Also of note in Fig. 4 is a slight positive bias in the *LPI* predictions, which stems from the fact *LPI* = 0 can only be overpredicted (such cases are more common at lower $PGA_M$, near the threshold for triggering). This bias averages $LPI = +1.0$ for $0 < PGA_M < 1$ g and is similar in the other ML models, as summarized in Table S4.

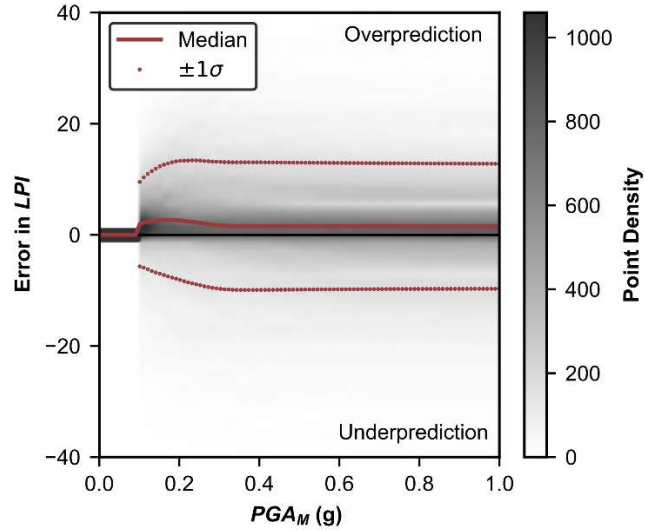

**Fig. 4.** *LPI* residuals as a function of $PGA_M$ for the global model test set.

This process was repeated for all global and New Zealand models, for which the results on the test set are summarized in Table 3 (complete performance statistics are provided in Table S4). The ML methodology's ability to mimic predicted *LPI* and $LPI_{ISH}$ values is very similar, whereas greater MAE and MSD are observed in the *LSN* predictions. However, because the mapping of *MI* to *PGF* is least sensitive for variations in *LSN* (e.g., using the fragility functions of Geyin and Maurer (2020)), the larger errors in predicted *LSN* do not usually translate to larger errors in *PGF*. Overall, the results in Table 3 suggest the ML models are similarly effective at mimicking *PGF* predictions based on any of the three geotechnical models. It must be emphasized, of course, that accurately mimicking the predictions of a geotechnical model does not guarantee accurate predictions of liquefaction phenomena (the ML models' abilities to predict liquefaction will be tested momentarily). Geotechnical models may also have different efficacies, albeit there is generally too little global liquefaction case-history data to establish statistical significance or consensus on which models perform best (e.g., Geyin et al., 2020; Rasanen et al., 2023). For this reason, users may wish to ensemble the predictions from one or more of the ML models developed here.

Table 3. Summary of test set performance for global and New Zealand models.

| Model | A | | B | | MI | | PGF | |
|---|---|---|---|---|---|---|---|---|
| | MAE | Standard Deviation | MAE | Standard Deviation | MAE | MSD | MAE | MSD |
| Global | | | | | | | | |
| $LPI$-ML | 3.0 | 7.0 | 5.0 | 15.5 | 4.5 | 11.3 | 8% | 22% |
| $LPI_{ISH}$-ML | 3.0 | 6.8 | 6.0 | 17.1 | 4.6 | 11.1 | 6% | 25% |
| $LSN$-ML | 4.0 | 10.5 | 18.0 | 26.8 | 4.9 | 16.7 | 7% | 22% |
| New Zealand | | | | | | | | |
| $LPI$-ML | 7.0 | 9.7 | 3.0 | 9.5 | 9.5 | 15.9 | 5% | 24% |
| $LPI_{ISH}$-ML | 7.0 | 9.9 | 3.0 | 10.4 | 9.8 | 16.5 | 4% | 25% |
| $LSN$-ML | 9.0 | 14.7 | 21.0 | 31.6 | 12.5 | 23.4 | 8% | 22% |

Although ML models have justly been criticized as opaque, interrogative techniques are continually advancing and the ability to understand ML predictions is nearing that of traditional regression. Insights can be gained, for example, from the computed predictor importance (e.g., Auret and Aldrich, 2011), which may be interpreted as each variable's relative contribution to model predictions. This method of ML interpretation has been used for prior geohazard models (e.g., Durante and Rathje, 2021; Geyin and Maurer, 2023). Because variable importances are similar across multiple models, we illustrate average importance for the global and New Zealand models in Fig. 5 (the 12 most important variables are shown) and provide complete results for all models in Figs. S50-S63. See Tables 1, S1, 2, and S2 for variable information.

Unsurprisingly, the most influential variable in both models is groundwater depth, which on average is ~300 times more influential than the least important variables included, one example of which is the 1-km resolution majority landform, which includes 10 classes (e.g., valley, footslope, ridge). At 90-m resolution, this variable is referred to as the "geomorphon" and becomes more important. Notably, it one of the more important variables in the New Zealand model. The global and New Zealand models share a handful of similarly important predictors, but the emergence of information derived from geology maps (e.g., "Simplified geology" and "Geology deposit type") as important in New Zealand substantiates the value of such knowledge. Near-surface geologic information, which is not consistently available at global scale, was substituted using proxy models such as the predicted "Sand fraction," but future iterations of the global model will look to benefit from improved geologic characterization. It should be noted that predictor importance describes a model's behavior, which does not necessarily reflect correlation between predictors and targets, independent of a model. A variable could conceivably correlate to a target, but if it also correlates with other variables, it may have diminished influence on that model's predictions.

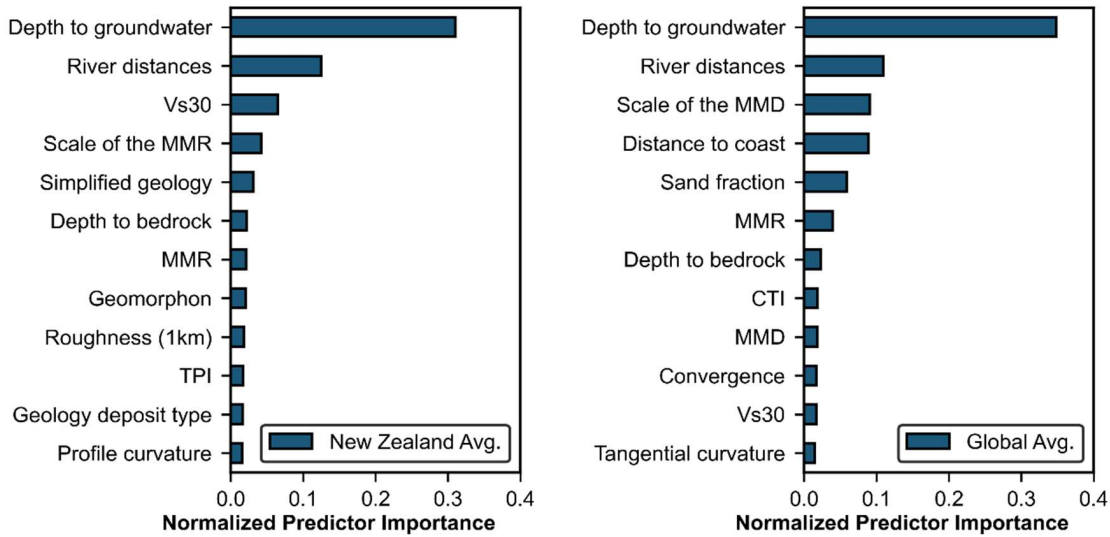

**Fig. 5.** Average normalized predictor importance for the (a) global and (b) New Zealand models.

The results in Fig. 5 allude to both the utility and insufficiency of variables in prior geospatial models. Three of the four capacity variables in Rashidian and Baise (2020), for example, are measured distance to water (i.e., coastline or rivers) and predicted groundwater depth and $V_{S30}$. These three variables are consistently important in the ML models. Yet other proxies for soil thickness (e.g., predicted depth to rock), saturation (e.g., height above nearest drainage, compound topographic index), depositional environment (e.g., maximum multiscale deviation, roughness), and typology (e.g., mapped surface geology, predicted sand and silt fractions) are also influential within the model architecture and lead to predictions that are spatially more nuanced. Notably, the distance to rivers has more gradation here than in other geospatial models in that it includes distances to seven different flow volumes (e.g., major rivers, but also seasonal drainages, are separately considered). The results in Fig. 5 have implications for forward application, since errors in the most influential variables (e.g., a mispredicted groundwater depth or surface geology) are more likely to propagate to errors in liquefaction predictions.

The trained models were implemented to predict *A* and *B* values worldwide by sampling variables at a resolution of ~90 m (0.000833 degrees). Substantial storage, memory, and processing capacity were required to: (i) sample ~40 global variables with total file size exceeding 1 TB; (ii) use these variables to make predictions with all ML models; and (iii) repeat for ~1.3 billion locations on earth. HPC resources were required to meet demands and were acquired via DesignSafe at the Texas Advanced Computing Center (Rathje et al., 2017) and the University of Washington. To minimize computational burdens and file sizes for end-users, and to reduce model extrapolations beyond the training data, predictions were made only for locations with 90-m topographic slope < 5° (Amatulli et al., 2020). This describes ~98% of CPT sample locations. Predictions were also not made for lakes (Messager et al., 2016), glaciers (RGI Consortium, 2023), the Greenland Ice Sheet (Lewis, 2009), and permafrost, both continuous and discontinuous (Brown et al., 2002). A few small and generally uninhabited islands were also excluded. With these exceptions, the global and New Zealand models have continuous coverage.

Predictions of *A* and *B* were next geostatistically updated in the vicinities of CPT measurements via regression kriging (e.g., Hengl et al., 2007), which merges model predictions (i.e., "regression") with spatial interpolation of model residuals (i.e., "kriging"). With this approach, *A* and *B* residuals are predicted using nearby CPTs (where residuals are known), and these predictions are used to update the *A* and *B* models as needed. Central to this approach is a semivariogram, which describes the spatial correlation of residuals. Having considered several common, monotonic semivariogram functional forms, including spherical, exponential, and Gaussian, a stable semivariogram was chosen for its best fit of residuals across all models:

$$Semivariance\ (h) = b + c_0 \left(1 - e^{-h^\alpha/r^\alpha}\right) \text{(Eq. 2)}$$

where $b$ is the nugget, or non-spatial variance; $c_0$ is the sill which describes the variance of residuals at distances beyond the range, where residuals become uncorrelated; $h$ is the separation distance between locations; $r$ is the effective range, or length scale of the model, which represents the distance over which correlation significantly decreases; and $\alpha$ is a shape parameter (Wackernagel, 2003). To propagate geotechnical adjustments on a scale consistent with the geospatial input information, the semivariance was computed using a 3-km neighborhood radius from each CPT and residual prediction was limited to 1.2 km (where the coarsest resolution of the geospatial predictors is approximately 1.1 km, see Table 1 and Table 2). Semivariograms were individually fitted to residuals for the 12 $A$ and $B$ models and the resulting parameters are summarized in Table S5. All semivariograms are provided in Figs. S64-S75. Using this information, residuals were spatially predicted for all ML models. Predicted residuals approach observed residuals at CPT sites and decay with distance toward zero (the mean residual for all models), governed by the semivariogram in Eq. (2). In parallel, the variance of residuals approaches zero at CPT sites and increases toward the overall model uncertainty at locations distant from CPTs. It should be noted that the nugget in Eq. (2), which governs residuals at a separation distance of zero (i.e., at CPT sites), is zero, meaning geotechnical measurement or model uncertainties are not considered. These could be the uncertainties of CPT measurements or those of $LPI$, $LPI_{ISH}$, and $LSN$. The nugget could also reflect sources of spatial variation at distances smaller than sampling intervals. In other words, $A$ and $B$ are unlikely to be constant over a 90-m map pixel, contrary to how the maps could be interpreted. However, because the nugget is not well constrained by the empirical data and would require judgement to define, it is here resigned to zero, which is a common default in kriging. This could be revisited in future model iterations.

Using kriged residuals, the global and New Zealand models were updated such that predictions of $A$ and $B$ (and by corollary, predictions of liquefaction response) are scaled up or down in the vicinity of CPTs, thereby anchoring the models to known conditions. To convey the degree to which predictions are updated by local geotechnical data, the variance of residuals modeled by regression kriging is given in an accompanying set of maps. We opt to map a classification of these variances: 3 = total ML model variance (i.e., no geotechnical influence), 2 = majority ML model variance (i.e., minor geotechnical influence); 1 = minority ML model variance (i.e., moderate geotechnical influence); and 0 = little to no ML model variance (i.e., major geotechnical influence). These maps thus communicate where, and to what degree, the predicted response is influenced by geotechnical data and models. To this end, Fig. 6 demonstrates an example of updating for the $LPI$ model and the associated variance classifications in San Bernadino, California.

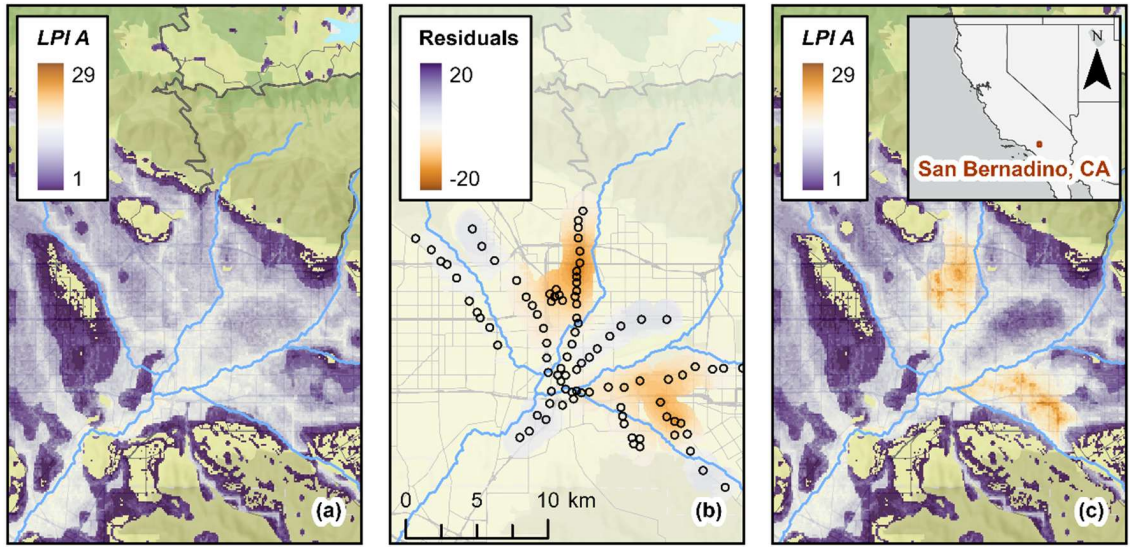

**Fig. 6.** An example in San Bernadino, California to illustrate *LPI A* a) before geotechnical updating, b) in terms of kriged residuals, and c) after geotechnical updating. (Sources: Esri, TomTom, FAO, NOAA, USGS, and CGIAR.)

The updated global and New Zealand models and the related variance classifications are provided as 90-m resolution geotiff files from Sanger et al. (2024b,c) (see *Data Availability*) with separate *A* and *B* rasters for each of the three geotechnical models. The global models are further parsed into seven geographical regions. One of these (Oceania) includes New Zealand, providing one example from which the benefits of region-specific GLMs can be judged. The complete global and New Zealand file packages are respectively 33 GB and 85 MB. However, if executing one global model (e.g., *LPI*) for one continent (e.g., North America), the required files diminish to ~1.5 GB. When combined with a "ShakeMap" of $PGA_M$, parameters *A* and *B* produce predictions of *LPI*, $LPI_{ISH}$, and *LSN* via Eq. (1). Because these predictions have, in effect, been made for all locations and all possible earthquakes, the expected liquefaction response is queried at very low computational expense. A simple script is provided by Sanger et al. (2024d) to implement any of the developed models in Python (Jupyter Notebook) and Matlab. These scripts are written to interact with USGS ShakeMaps in .xml format, which are called via a user-input web address. ShakeMaps are easily obtained from USGS or analogous global organizations, both for countless scenario earthquakes and for those that have just occurred. As with other geospatial models, predictions can be made in near-real-time to inform response, reconnaissance, and decision-making in the aftermath of an event. For consistency with how the models were trained, $PGA_M$ should be computed from $PGA$ with the magnitude-scaling factor of Idriss and Boulanger (2008):

$$PGA_M = \frac{PGA}{MSF}, where\ MSF = 6.9\ exp\left(\frac{-M}{4}\right) - 0.058 \leq 1.8 \qquad (Eq.\ 3)$$

where M = moment magnitude and *PGA* is that at the surface, having been corrected for site effects (e.g., the *PGA* in any USGS ShakeMap). The resulting event-specific mapped predictions of *LPI*, $LPI_{ISH}$, and *LSN* can be propagated via fragility functions, or "damage" functions, that have been conditioned on *MI* to predict the probabilities of various outcomes (e.g., Geyin and Maurer, 2020, Toprak et al., 2019, Maurer et al., 2025). In this paper, results are presented as probability of liquefaction-induced ground deformation or ejecta observed at the surface (i.e., the median probability of observing liquefaction manifestation at any location within the given map pixel, "*PGF*") using the fragility functions of Geyin and Maurer (2020), which we recommended for general use with the proposed models. Of added utility, Geyin and Maurer (2020) considered finite-sample uncertainty and provided $16^{th}$, $50^{th}$, and $84^{th}$ percentile fits for their functions. This permits the uncertainty of the *PGF*, conditioned on the *MI* value, to be quantified. Although only the median functions are used here, an example including this uncertainty is given in Fig. S76.

To demonstrate model application and the effects of updating, the global model is here applied to the 22 February 2011, $M_W$6.2 Christchurch, New Zealand, earthquake. Results are shown in Fig. 7 for the Burwood neighborhood, which experienced widespread liquefaction. CPT sites are also mapped and are symbolized based on whether liquefaction manifestations were observed, as compiled by Geyin et al. (2021). Predictions by the Rashidian and Baise (2020) geospatial model, henceforth RB20, are shown in Fig. 7A and somewhat underpredict manifestations, with sites of positive observation having a modal probability of 47%. Predictions by the global ML model, before and after updating, are shown in Figs. 7B and 7C, respectively. As compared to RB20, the ML model's predictions tend to be higher, especially in the east of the mapped area, and have more spatial nuance due to the inclusion of more geospatial information. This nuance is increased by updating, which in Fig. 7C can generally be observed to improve predictions. ML predictions tend to be scaled up and down, respectively, in areas with and without observed liquefaction.

It is emphasized that this updating is not driven by liquefaction observations, but rather, by geotechnical data and models that more correctly predict these observations. The classified variance of kriged residuals is shown in Fig. 7D, from which a user can quickly understand where predictions are predominated by geotechnical models, and where they are purely those of ML.

The ML models developed here will ultimately be judged in the context of predicting liquefaction in the field. To that end, we conduct tests to answer three research questions and compare predictions against RB20. Because variants of RB20 are in use, we also execute that which is adopted in the USGS ground failure product with ad-hoc modifications (Allstadt et al., 2022), as well as the Zhu et al. (2017) model upon which RB20 is based. These three versions are very similar; thus, we report performance only for that which performs best in each test. To quantify model performance, the Brier Score ($BS$) is adopted:

$$\text{Brier Score } (BS) = \frac{1}{N}\sum_{i=1}^{N}(P_i - O_i)^2 \quad \quad \text{(Eq. 4)}$$

where $P$ is the predicted probability, $O$ is the observed probability (0 or 1), $N$ is the number of observations, and $i$ is the observation index. The $BS$ is essentially MSE for probabilistic classification models. $BS = 0$ defines a perfect model, $BS = 0.5$ represents a model that randomly predicts the outcome, and $BS = 0.25$ represents a model that predicts a probability of 50% for every event. Therefore, a $BS < 0.25$ is considered a "good" model, and increasingly good as $BS$ approaches 0. The $BS$ simultaneously measures: (i) the degree to which positive and negative class distributions are segregated by a model; and (ii) the degree to which this segregation centers on a probability of 50%. Although the first of these could instead be measured by the area under a receiver-operating-characteristic curve (i.e., ROC $AUC$) (e.g., Fawcett, 2006), $AUC$ does not consider the latter, which is a critical metric of model portability across events. A model could have a perfect $AUC$ but the model output at which two classes are best separated could be very far from 50% probability. As an alternative graphical representation to the ROC, a calibration curve, or reliability diagram, is adopted. In these diagrams, a perfectly calibrated model follows the 1:1 line, whereas models that overpredict or underpredict response respectively plot below, and above, the 1:1 line. Together, the *BS* and calibration curve describe performance in general terms of magnitude and direction of mispredictions, respectively.

To account for finite-sample uncertainty and assess statistical significance, we bootstrap the *BS* test results 10,000 times to compute *BS* confidence intervals (CIs), perform the Kolmogorov-Smirnov (KS) two-sample test (Smirnov, 1939), and compute Cohen's *d* effect (Cohen, 1988). These outputs serve as metrics to determine whether the computed differences in model performance are statistically significant or could have arisen by chance. To account for the possibility of spatial correlation in the *BS* data, which if present might obfuscate model comparisons, we compute the Moran's *I* statistic (Anselin, 1996) to measure autocorrelation. When this statistic exceeds 0.3, the *BS* distribution is resampled with spatial stratification via agglomerative clustering (e.g., Steinbach, 2000), which intends to remove spatial correlation. Ultimately, because the *BS* data are only weakly correlated in most of the subsequent tests, the test metrics are only minimally altered and the conclusions drawn from these tests are insensitive to the resampling. The two-sample KS test is a nonparametric test used to determine whether two distributions are significantly different. It quantifies the maximum absolute difference between two cumulative distribution functions, where a KS statistic between 0 and 0.2 suggests little to no difference, values between 0.2 and 0.5 indicate moderate differences, and values greater than 0.5 indicate the distributions are significantly different. Cohen's *d* effect provides a metric for comparing two distributions as the difference in their means normalized by the pooled standard deviation. This quantifies both the magnitude and direction of the difference between two groups, or distributions. Cohen's *d* effects below 0.2 suggest little to no difference, values between 0.2 and 0.5 indicate a small difference, values between 0.5 and 0.8 represent a moderate difference, and values greater than 0.8 indicate a large difference (a *d* of 1 indicates the groups differ by one standard deviation). The sign of Cohen's *d* denotes the direction of the difference, where a negative value indicates that the *BS* of the treatment group (i.e., the ML model) is less than that of the control group (i.e., the RB20 model).

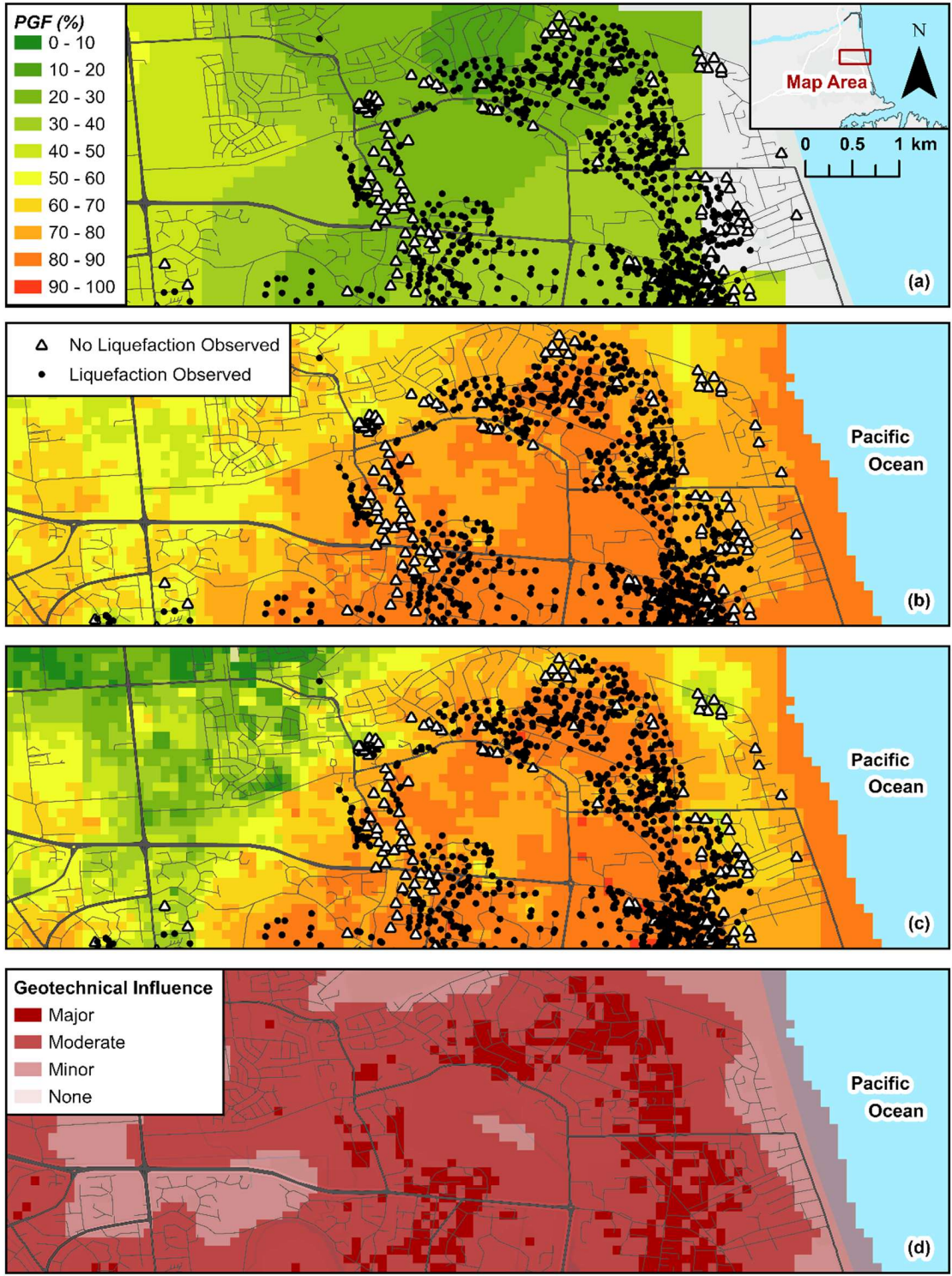

**Fig. 7.** An example to illustrate the predicted *PGF* for the 22 February 2011, Mw6.2 Christchurch, New Zealand, earthquake event according to a) RB20, and the global *LPI* model b) before and c) after geotechnical updating, as well as the d) classified variance of geotechnical model influence. (Sources: Stats NZ, Esri, TomTom, Garmin, FAO, NOAA, USGS, LINZ, and METI/NASA.)

*Testing performance on "unseen" case histories*

The first research question is: how does the ML model perform prior to updating with geotechnical data and models? In other words, how do the ML and RB20 models compare in regions unknown to each model's training set and devoid of CPT data? We use three liquefaction inventories that postdate RB20's training set and which occurred in regions where no CPTs were compiled in the current effort: the 2019 Ridgecrest (Zimmaro et al., 2020), 2019 Puerto Rico (Allstadt and Thompson, 2021), and 2023 Turkey earthquakes (Cetin et al., 2023; Taftsoglou et al., 2023). Negative observations were randomly sampled from the extents of each event's ShakeMap, such that map cells without positive observations were assumed negative. Although this assignment is obviously uncertain, it is a pragmatic and common assumption in the geospatial modeling literature, permitting an assessment of performance to be made over a very large area. In addition to the three surrogate geotechnical models, we test the performance of these models when averaged, or ensembled. The results of this test are in Table 4. The $LPI_{ISH}$ model had a mean $BS$ of 0.128 and performed best of the three ML models, which all outperformed RB20 (mean $BS$ = 0.393) to an apparently significant degree, as indicated by 99% CIs and KS test statistics on the $BS$ distributions. Cohen's $d$ indicates a very large negative effect, such that the ML models reduced the $BS$ of RB20 by about 1.5 standard deviations. The calibration curves in Fig. S77 indicate the ML models had some tendency to underpredict liquefaction response for these events, whereas RB20 underpredicted this response even more severely.

**Table 4.** Summary of global model performance in unbiased testing.

| Model | Mean $BS$ | 99% Confidence Interval of Mean $BS$ | Comparison Against RB20 | |
|---|---|---|---|---|
| | | | KS Test Statistic | Cohen's $d$ Effect |
| RB20 | 0.393 | 0.380 - 0.407 | - | - |
| $LPI$-ML | 0.153 | 0.143 - 0.162 | 0.56 | -1.51 |
| $LPI_{ISH}$-ML | 0.128 | 0.120 - 0.137 | 0.62 | -1.62 |
| $LSN$-ML | 0.180 | 0.170 - 0.191 | 0.49 | -1.46 |
| Ensemble | 0.146 | 0.138 - 0.155 | 0.57 | -1.59 |

*Testing distributed global performance before and after updating*

The second question is: does updating improve model performance? In other words, how does each ML model perform before and after updating at sites with CPTs, and how does this compare to RB20? We adopt the inventory of 332 liquefaction case histories compiled from 25 global earthquakes by Rateria et al. (2024). This compilation includes both positive and negative observations made at the locations of CPTs. This evaluation includes bias that is difficult to quantify. RB20 previously trained on liquefaction inventories from 21 of these earthquakes and the ML models were similarly trained on CPTs from ~90% of sites tested here (albeit these sites represent less than 1% of the total training set). In these tests the three ML models performed very similarly and showed similar improvements from updating, as summarized in Table 5. The $LPI_{ISH}$ ML model again performed best, with mean $BS$ of 0.209. Relative to both the preceding and ensuing tests, the ML models are somewhat less efficacious. This is not unexpected, given that this dataset samples a broad array of conditions and environments across 25 events and considering that the negative case histories comprising it are both sampled from susceptible deposits and proximal to positive observations. Because our geospatial variables have 90-m and coarser resolution, they are ill equipped to predict variations in subsurface response across finer spatial scales. The 99% CIs suggest all three ML models outperformed RB20 by a statistically significant degree, especially after updating. The KS test statistics indicate that the ML and RB20 distributions are moderately different, whereas the large negative Cohen's $d$ indicates the ML models reduced the $BS$ of RB20 by about one standard deviation. The calibration curves in Fig. S78 indicate that the ML models generally slalom about the 1:1, or perfect calibration, line both before and after updating, whereas RB20 exhibits stronger patterns of both under- and overprediction.

Table 5. Summary of global model performance in global case histories.

| Model | Mean BS | 99% Confidence Interval of Mean BS | Comparison Against RB20 | |
|---|---|---|---|---|
| | | | KS Test Statistic | Cohen's $d$ Effect |
| RB20 | 0.299 | 0.292-0.305 | - | - |
| **Before Updating** | | | | |
| $LPI$-ML | 0.231 | 0.226-0.237 | 0.24 | -1.05 |
| $LPI_{ISH}$-ML | 0.228 | 0.223-0.233 | 0.22 | -1.11 |
| $LSN$-ML | 0.234 | 0.228-0.241 | 0.19 | -0.99 |
| Ensemble | 0.228 | 0.222-0.233 | 0.24 | -1.09 |
| **After Updating** | | | | |
| $LPI$-ML | 0.226 | 0.220-0.232 | 0.25 | -1.03 |
| $LPI_{ISH}$-ML | 0.209 | 0.205-0.214 | 0.23 | -1.19 |
| $LSN$-ML | 0.226 | 0.220-0.233 | 0.26 | -0.96 |
| Ensemble | 0.208 | 0.203-0.213 | 0.25 | -1.10 |

*Testing the efficacy of regional models*

The third question is: can model regionalization improve performance? We adopt an inventory of 16,836 observations compiled by Geyin et al. (2021) after three earthquakes in Canterbury, New Zealand: the 4 Sept. 2010 $M_w$7.1 Darfield, 22 Feb. 2011 $M_w$6.2 Christchurch, and 14 Feb. 2016 Christchurch ruptures. Because these observations were made at CPT sites, the global and New Zealand models give nearly identical predictions after geotechnical updating. For this reason, we assess ML model performance prior to updating and compare against RB20, as summarized in Table 6. In general, the region-specific ML models perform marginally better than their global counterparts. The less than dramatic improvement suggests that region-specific models may not be successful elsewhere, given that New Zealand has both considerable geotechnical data and high-quality regional variables. These advantages may be outweighed by the benefits of learning from substantially more global data. The $LPI_{ISH}$ model, with mean $BS$ of 0.127 (global) and 0.134 (New Zealand), was again the best of the ML models. The 99% CIs suggest the ML models all give statistically significant improvements over RB20, which also performed relatively well on this dataset, with mean $BS$ of 0.204. KS statistics suggest these improvements are moderate to significant, depending on which ML model is chosen, and whether it is the global or New Zealand variant. Cohen's $d$ effect indicates the improvements are large and in the range of 1.1-1.3 standard deviations. The calibration curves in Fig. S79 indicate that both RB20 and the ML models track the ideal 1:1 line relatively well, but with some slight tendency toward underprediction, especially when the predicted probabilities exceed 0.5. Notably, these tests are not without bias. RB20 trained on inventories from two of these three events, which all affected the same area, and the ML models were trained on CPTs from this affected region, albeit these CPTs were down weighted during training due to their high spatial density. Collectively, results from the three tests suggest that the ML models developed herein warrant adoption and further testing.

**Table 6.** Summary of global and New Zealand model performance in Canterbury case histories.

| Model | Mean BS | 99% Confidence Interval of Mean BS | Comparison Against RB20 | |
|---|---|---|---|---|
| | | | KS Test Statistic | Cohen's $d$ Effect |
| RB20[a] | 0.204 | 0.200-0.207 | - | - |
| Global | | | | |
| $LPI$-ML[b] | 0.143 | 0.139-0.147 | 0.40 | -1.20 |
| $LPI_{ISH}$-ML | 0.127 | 0.123-0.132 | 0.52 | -1.08 |
| $LSN$-ML | 0.187 | 0.183-0.191 | 0.21 | -1.30 |
| Ensemble | 0.146 | 0.142-0.149 | 0.40 | -1.24 |
| New Zealand | | | | |
| $LPI$-ML | 0.142 | 0.138-0.146 | 0.42 | -1.17 |
| $LPI_{ISH}$-ML | 0.134 | 0.129-0.138 | 0.45 | -1.09 |
| $LSN$-ML | 0.161 | 0.156-0.166 | 0.37 | -1.16 |
| Ensemble | 0.140 | 0.136-0.144 | 0.41 | -1.18 |

[a,b] These *PGFs* are mapped in (a) Fig. 7A and (b) Fig. 7B for the Burwood neighborhood of Christchurch for the 22 February 2011, $M_w$6.2 earthquake (i.e., one of three events in this case history set).

*Limitations, Uncertainties, and Future Work*

The developed models are subject to limitations and uncertainties not yet discussed. *First*, using triggering models other than Idriss and Boulanger (2008) could have altered predictions of *LPI*, *LPI*$_{ISH}$, and *LSN*. However, because the fragility functions used in forward application are specific to Idriss and Boulanger (2008), any systematic shifts in triggering predictions by another model would be mitigated using a fragility function specific to that model (e.g., Geyin and Maurer, 2020). More broadly, the adopted geotechnical models are imperfect and will inevitably be supplanted. Recent tests of these models on global case histories demonstrate an approximate median AUC of 0.78 (Geyin et al., 2020). It is known, for example, that these models are less constrained in certain domains of their parameter spaces (e.g., there are fewer training data from $M_W > 8$ events) and that they may perform less efficiently on highly stratified profiles (e.g., Cubrinovski et al., 2019; Geyin and Maurer, 2021b). New triggering and manifestation models are frequently proposed in the attempt to address these and other limitations (e.g., Hutabarat and Bray, 2022; Upadhyaya et al., 2023). Our methodology should therefore improve as the underpinning geotechnical models improve, and as additional geotechnical data become available, both for training and updating. An important caveat pertains to lateral spreading. Although cases of lateral spreading were included in the preceding tests, they depend on factors not considered by $LPI$, $LPI_{ISH}$, nor $LSN$, which can thus predict it poorly (e.g., Maurer et al. 2015b). This might be improved by merging the predicted $LPI$ with topographic data, as formulated by Rashidian and Gillins (2018), but this possibility was not tested.

*Second*, CPTs may be preferentially performed in ground where liquefaction hazards are expected and/or where premature refusal is less likely. If so, the proposed models might overpredict liquefaction, particularly in regions unrepresented in training, albeit the limited tests performed here do not confirm this expectation. Nonetheless, SPTs could help evaluate this possibility and improve the model in geologies and regions where CPTs are uncommon. Moreover, it should be recognized that model uncertainties could exceed those indicated by the test statistics in data-poor areas. We partially assess this possibility in Fig. S80, where prediction residuals for $LPI$ $A$ and $B$ are binned based on the spatial density of CPTs. The residual means are ~zero, and thus unbiased, across all spatial densities, and there is no clear trend in the spread of the residuals with density. As more geotechnical data become available, the presented models can be updated in two ways: (i) model retraining; and (ii) model re-kriging. The first is computationally expensive and unlikely to result in major changes unless the new data are plentiful or otherwise expand the parameter space of the current training set. The second could feasibly be done frequently, including by those who wish to geostatistically update the ML models with emergent or proprietary data for a specific municipality, large project site, or network of distributed infrastructure.

*Third*, the ML models are inherently limited by the accuracies and spatial resolutions of geospatial predictors, some of which are themselves prediction models. Mispredictions of liquefaction are therefore more likely where subsurface conditions change at a finer scale than the geospatial variables (e.g., as indicated by the high variance observed in semivariograms for some $B$ models, particularly at scales finer than the geospatial variables) or where subsurface conditions are otherwise uncaptured by those variables (e.g., where influential and potentially dynamic variables such as the predicted groundwater depth are inaccurate). Terrain abutting flat land could erroneously be predicted to liquefy, for example, or deposits highly susceptible to liquefaction, such as artificial fill, could go unnoticed unless sampled by CPTs. It is also conceivable that variables could be judged as unimportant in the current models, or may have been omitted entirely, because the training data are insufficient to elucidate their predictive value. Conversely, variables could mistakenly be judged as important if correlations in the data falsely suggest causality. This is true of any empirical model. Although domain knowledge was used to omit variables and several overfitting techniques were employed, future versions of this model will inevitably benefit from additional, higher resolution data and will explore groundwater depth as a user input variable á la $PGA_M$.

*Fourth*, it should be noted that while several aspects and sources of uncertainty have been discussed, we do not attempt a rigorous accounting of all uncertainties. The Geyin and Maurer (2020) fragility functions allow for consideration of one salient uncertainty: that of the $PGF$ conditioned on an $MI$ (e.g., $LPI$), but this is not a panacea for all sources of uncertainty entering the analyses. Lastly, geostatistical updating could be performed using other methods that could alter expectations of liquefaction in the vicinity of CPTs. Our updating was not explicitly bound by predictor variables but possibly could be. As an example, an overprediction of *A* or *B* in a sandy deposit may not necessarily indicate that *A* or *B* is also overpredicted in a gravelly deposit several hundred meters away, in contrast to what a univariate semivariogram conveys. Although improvements are inevitably warranted, this study proposed and demonstrated a new approach to developing GLMs that arguably has many merits. Ultimately, additional data and analyses will verify or refine the results shared here and succinctly summarized below.

**User Implementation**

With proper consideration of their limitations and uncertainties, the proposed models may be used to preliminarily assess liquefaction hazards for conventional geotechnical engineering projects, where the prediction is a general, site-averaged hazard. However, the ultimate utility of our approach rests in regional-scale applications, both for scenario events and for earthquakes that have just occurred. In this regard, the models can be used to predict impacts on linear infrastructures and other spatially distributed assets (e.g., utilities or transportation systems), and to inform response, reconnaissance, evacuation routes, and other planning needs immediately after an event.

Example model-use scripts (i.e., a Jupyter Notebook and Matlab script) are provided by Sanger et al. (2024d), wherein the user defines the geospatial model of interest (i.e., the global or New Zealand specific model), the surrogate geotechnical model of interest (i.e., $LPI$, $LPI_{ISH}$, or $LSN$), and an input ground motion via a USGS Shakemap URL. The scripts are configured to run in the DesignSafe environment and reference the model datasets published in DesignSafe (i.e., Sanger et al., 2024b,c). For most earthquakes, the DesignSafe virtual machine environment is sufficient. For events impacting extremely large areas, such as an $M_W9$ rupture of the Cascadia Subduction Zone, the scripts may need to be executed in a DesignSafe HPC instance with more memory resources.

The example code produces four deliverables: (i) a geotiff of $PGA$, interpolated to the resolution of the GLM; (ii) a geotiff of $PGA_M$; (iii) a geotiff of the predicted liquefaction response in terms of the selected surrogate model; and (iv) a geotiff of the predicted probability of ground failure (i.e., surface deformation and ejecta) according to the fragility functions of Geyin and Maurer (2020). Additional guidance is available in the ReadMe file of Sanger et al. (2024d). Alternatively, the models may be executed in the GIS environment using Eqs. 1 and 3 in conjunction with mapped parameters $A$ and $B$, at which point the $LPI$, $LPI_{ISH}$, or $LSN$ predictions can be paired with a fragility function of the user's choosing.

**Conclusions**

Using mechanics-informed machine learning, this study developed surrogate models to predict soil liquefaction using geospatial information. Two models were trained to mimic three geotechnical models: one globally applicable, and one specific to New Zealand. These models have conceptual advantages over prior geospatial approaches and were shown to provide improved predictions in test applications. These tests suggested that the geospatial ML models themselves (i.e., prior to geotechnical updating) outperform existing models, and that updating further improves their performance. Tests of the New Zealand model suggested that while region-specific models may perform as well or better than their global complements, their benefits could largely be negated by the advantages of learning from substantially larger global datasets. Although developed using a large body of geospatial data, machine learning, and HPC, the models are packaged in a practical format that requires only simple arithmetic to execute, and which encourages user adoption and testing.

**Data Availability**

The geotechnical and geospatial data used in model development are all publicly available, as described and referenced in the text. The model products are available on DesignSafe, including: (i) global GLM geotiffs for *LPI*, *LPI$_{ISH}$*, and *LSN* (Sanger et al., 2024b); and (ii) New Zealand GLM geotiffs for *LPI*, *LPI$_{ISH}$*, and *LSN* (Sanger et al., 2024c).

**Acknowledgements**

The presented work is based on research supported by the United States Geological Survey (USGS) under award G23AP00017, the Cascadia Region Earthquake Science Center (CRESCENT) via National Science Foundation (NSF) award 2225286, the Cascadia Coastlines and Peoples (CoPes) Hub via NSF award 2103713, the Pacific Earthquake Engineering Research (PEER) Center under award 1185-NCTRMB, and the Pacific Northwest Transportation Consortium (PacTrans) under award 69A3552348310. However, any opinions, findings, conclusions, or recommendations expressed herein are those of the authors and may not reflect the views of USGS, NSF, CoPes, CRESCENT, PEER, or PacTrans. Furthermore, this work was facilitated using HPC infrastructure provided by the Hyak supercomputer and funded by the University of Washington's student technology fund, and through DesignSafe at the Texas Advanced Computing Center.

**Supplemental Materials**

Figs. S1–S80 and Tables S1-S5 are available online in the ASCE Library ([www.ascelibrary.org](www.ascelibrary.org)).

**References**

Allstadt, K.E., & Thompson, E.M. (2021). *Inventory of liquefaction features triggered by the 7 January 2020 M6.4 Puerto Rico earthquake* [Dataset]. USGS. https://doi.org/10.5066/P9HZRXI9
Allstadt, K. E., Thompson, E. M., Jibson, R. W., Wald, D. J., Hearne, M., Hunter, E. J., Fee, J., Schovanec, H., Slosky, D., & Haynie, K. L. (2022). The US Geological Survey ground failure product: Near-real-time estimates of earthquake-triggered landslides and liquefaction. *Earthquake Spectra*, *38*(1), 5-36. https://doi.org/10.1177/87552930211032685

Amatulli, G., Domisch, S., Tuanmu, M. N., Parmentier, B., Ranipeta, A., Malczyk, J., & Jetz, W. (2018). A suite of global, cross-scale topographic variables for environmental and biodiversity modeling. *Scientific data*, *5*(1), 1-15. https://doi.org/10.1038/sdata.2018.40

Amatulli, G., McInerney, D., Sethi, T., Strobl, P., & Domisch, S. (2020). Geomorpho90m, empirical evaluation and accuracy assessment of global high-resolution geomorphometric layers. *Scientific Data, 7*(162). https://doi.org/10.1038/s41597-020-0479-6

Anselin, L. (1995). The Local Indicators of Spatial Association – LISA. *Geographical Analysis*, 27, 93-115. https://doi.org/10.1111/j.1538-4632.1995.tb00338.x

Asadi, A., Sanon, C., Cakir, E., Zhan, W., Shirzadi, H., Baise, L. G., Cetin, K. O., & Moaveni, B. (2024). Geospatial liquefaction modeling of the 2023 Türkiye earthquake sequence by an ensemble of global, continental, regional, and event-specific models. *Seismological Research Letters*, *95*(2A), 697-719. https://doi.org/10.1785/0220230287

Auret, L., & Aldrich, C. (2011). Empirical comparison of tree ensemble variable importance measures. *Chemometrics and Intelligent Laboratory Systems*, *105*(2), 157-170. https://doi.org/10.1016/j.chemolab.2010.12.004

Azul, K. M., Orense, R., & Wotherspoon, L. (2024). Investigation of representative geotechnical data for the development of a hybrid geotechnical-geospatial liquefaction assessment model. *Japanese Geotechnical Society Special Publication*, *10*(17), 585-590. https://doi.org/10.3208/jgssp.v10.OS-6-07

Baker, J. W., Crowley, H., Wald, D., Rathje, E., Au, S. K., Bradley, B. A., Burton, H., Cabas, A., Cattari, S., Cauzzi, C., Cavalieri, F., Contereras, S., Costa, R., Eguchi, R. T., Lallemant, D., Lignos, D. G., Maurer, B. W., Hutt, C. M., Sextos, A., ... & Thompson, E. M. (2024). Sharing data and code facilitates reproducible and impactful research. *Earthquake Spectra, 40*(3), 2210-2218. https://doi.org/10.1177/87552930241259397

Boulanger, R. W., & DeJong, J. T. (2018). Inverse filtering procedure to correct cone penetration data for thin-layer and transition effects. In *Cone Penetration Testing 2018* (pp. 25-44). https://doi.org/10.1201/9780429505980

Boulanger, R. W., & Idriss, I. M. (2006). Liquefaction susceptibility criteria for silts and clays. *Journal of geotechnical and geoenvironmental engineering*, *132*(11), 1413-1426. https://doi.org/10.1061/(ASCE)1090-0241(2006)132:11(1413)

Boulanger, R. W., & Idriss, I. M. (2016). CPT-based liquefaction triggering procedure. *Journal of Geotechnical and Geoenvironmental Engineering*, *142*(2), 04015065. https://doi.org/10.1061/(ASCE)GT.1943-5606.0001388

Bozzoni, F., Bonì, R., Conca, D., Meisina, C., Lai, C. G., & Zuccolo, E. (2021). A geospatial approach for mapping the earthquake-induced liquefaction risk at the European scale. *Geosciences*, *11*(1), 32. https://doi.org/10.3390/geosciences11010032

Breiman, L. (1996), Bagging predictor. *Machine Learning, 24*(2), 123-140. https://doi.org/10.1007/BF00058655

Breiman, L. (1999). Pasting small votes for classification in large databases and on-line. *Machine Learning* 36, 85–103 https://doi.org/10.1023/A:1007563306331

Breiman, L. (2001). Random forests. *Machine Learning*, *45*(1), 5–32. https://doi.org/10.1023/A:1010933404324

Brown, J., Ferrians, O., Heginbottom, J. A., & Melnikov, E. (2002). *Circum-Arctic map of permafrost and ground-ice conditions, Version 2* [Dataset]. NASA National Snow and Ice Data Center Distributed Active Archive Center. https://doi.org/10.7265/skbgkf16

Bullock, Z., Zimmaro, P., Lavrentiadis, G., Wang, P., Ojomo, O., Asimaki, D., Rathje, E. M., & Stewart, J. P. (2023). A latent Gaussian process model for the spatial distribution of liquefaction manifestation. *Earthquake Spectra*, *39*(2), 1189-1213. https://doi.org/10.1177/87552930231163894

Cetin, K. O., Soylemez, B., Guzel, H., & Cakir, E. (2024). Soil liquefaction sites following the February 6, 2023, Kahramanmaraş-Türkiye earthquake sequence. *Bulletin of Earthquake Engineering*, 1-24. https://doi.org/10.1007/s10518-024-01875-3

Chen, T., & Guestrin, C. (2016). XGBoost: A Scalable Tree Boosting System. *Proceedings of the 22nd ACM SIGKDD International Conference on Knowledge Discovery and Data Mining*, 785–794. https://doi.org/10.1145/2939672.2939785

Cohen, J. (1988). *Statistical power analysis for the behavioral sciences*. Hillsdale, NJ: Lawrence Erlbaum.

Cubrinovski, M., Rhodes, A., Ntritsos, N., & Van Ballegooy, S. (2019). System response of liquefiable deposits. *Soil Dynamics and Earthquake Engineering*, *124*, 212-229. https://doi.org/10.1016/j.soildyn.2018.05.013

de Magistris, F. S., Lanzano, G., Forte, G., & Fabbrocino, G. (2013). A database for PGA threshold in liquefaction occurrence. *Soil Dynamics and Earthquake Engineering*, *54*, 17-19. https://doi.org/10.1016/j.soildyn.2013.07.011

Durante, M. G., & Rathje, E. M. (2021). An exploration of the use of machine learning to predict lateral spreading. *Earthquake Spectra*, *37*(4), 2288-2314. https://doi.org/10.1177/87552930211004613

Fan, Y., Li, H., & Miguez-Macho, G. (2013, updated 2019). Global patterns of groundwater table depth. *Science, 339*, 940-943. https://www.science.org/doi/10.1126/science.1229881

Fawcett, T. (2006). An introduction to ROC analysis. *Pattern recognition letters, 27*(8), 861-874. https://doi.org/10.1016/j.patrec.2005.10.010

Foster, K. M., Bradley, B. A., McGann, C. R., & Wotherspoon, L. M. (2019). A Vs30 map for New Zealand based on geologic and terrain proxy variables and field measurements. *Earthquake Spectra, 35*(4), 1865-1897. https://doi.org/10.1193/121118EQS281M

Geyin, M., & Maurer, B. W. (2020). Fragility functions for liquefaction-induced ground failure. *Journal of Geotechnical and Geoenvironmental Engineering*, *146*(12), 04020142. https://doi.org/10.1061/(ASCE)GT.1943-5606.0002416

Geyin, M., & Maurer, B. W. (2021a). *CPT-Based Liquefaction Case Histories from Global Earthquakes: A Digital Dataset (Version 1)* [Dataset]. DesignSafe-CI. https://doi.org/10.17603/ds2-wftt-mv37

Geyin, M., & Maurer, B. W. (2021b). Evaluation of a cone penetration test thin-layer correction procedure in the context of global liquefaction model performance. *Engineering Geology*, *291*, 106221. https://doi.org/10.1016/j.enggeo.2021.106221

Geyin, M., & Maurer, B. W. (2023). US national $V_{S30}$ models and maps informed by remote sensing and machine learning. *Seismological Society of America*, *94*(3), 1467-1477. https://doi.org/10.31224/2466

Geyin, M., Baird, A.J. & Maurer, B.W. (2020). Field assessment of liquefaction prediction models based on geotechnical vs. geospatial data, with lessons for each. *Earthquake Spectra, 36*(3), 1386–1411. https://doi.org/10.1177/875529301989995

Geyin, M., Maurer, B. W., Bradley, B. A., Green, R. A., & van Ballegooy, S. (2021). CPT-based liquefaction case histories compiled from three earthquakes in Canterbury, New Zealand. *Earthquake Spectra*, *37*(4), 2920-2945. https://doi.org/10.1177/8755293021996367

Geyin, M., Maurer, B. W., & Christofferson, K. (2022). An AI driven, mechanistically grounded geospatial liquefaction model for rapid response and scenario planning. *Soil Dynamics and Earthquake Engineering*, *159*, 107348. https://doi.org/10.1016/j.soildyn.2022.107348

Heath, D., Wald, D. J., Worden, C. B., Thompson, E. M., & Scmocyk, G. (2020). A global hybrid Vs30 map with a topographic-slope-based default and regional map insets. *Earthquake Spectra, 36*(3), 1570-1584.

Hengl, T. (2018a). *Soil bulk density (fine earth) 10 x kg / m-cubic at 6 standard depths (0, 10, 30, 60, 100 and 200 cm) at 250 m resolution (v0.2)* [Dataset]. Zenodo. https://doi.org/10.5281/zenodo.2525665

Hengl, T. (2018b). *Clay content in % (kg / kg) at 6 standard depths (0, 10, 30, 60, 100 and 200 cm) at 250 m resolution (v0.2)* [Dataset]. Zenodo. https://doi.org/10.5281/zenodo.2525663

Hengl, T. (2018c). *Sand content in % (kg / kg) at 6 standard depths (0, 10, 30, 60, 100 and 200 cm) at 250 m resolution (v0.2)* [Dataset]. Zenodo. https://doi.org/10.5281/zenodo.2525662

Hengl, T. (2018d). *Silt content in % (kg / kg) at 6 standard depths (0, 10, 30, 60, 100 and 200 cm) at 250 m resolution (v0.2)* [Dataset]. Zenodo. https://doi.org/10.5281/zenodo.2525676

Hengl, T., & Gupta, S. (2019). *Soil water content (volumetric %) for 33kPa and 1500kPa suctions predicted at 6 standard depths (0, 10, 30, 60, 100 and 200 cm) at 250 m resolution (v0.1)* [Dataset]. Zenodo. https://doi.org/10.5281/zenodo.2784001

Hengl, T., & Nauman, T. (2018). Predicted USDA soil great groups at 250 m (probabilities) (v0.2) [Data set]. Zenodo. https://doi.org/10.5281/zenodo.352806

Hengl, T., Heuvelink, G. B., & Rossiter, D. G. (2007). About regression-kriging: From equations to case studies. *Computers & geosciences*, *33*(10), 1301-1315. https://doi.org/10.1016/j.cageo.2007.05.001

Hutabarat, D., & Bray, J. D. (2022). Estimating the severity of liquefaction ejecta using the cone penetration test. *JGGE*, *148*(3), 04021195. https://doi.org/10.1061/(ASCE)GT.1943-5606.0002744

Idriss, I. M., & Boulanger, R. W. (2008). *Soil liquefaction during earthquakes*. Earthquake Engineering Research Institute.

Islam, M., Chen, G., & Jin, S. (2019). An Overview of Neural Network. *American Journal of Neural Networks and Applications*, *5*(1). https://doi.org/10.11648/j.ajnna.20190501.12

Iwasaki, T. (1978). A practical method for assessing soil liquefaction potential based on case studies at various sites in Japan. In *Proc. of 2nd Int. National Conf. on Microzonation, 1978* (Vol. 2, pp. 885-896).

Jena, R., Pradhan, B., Almazroui, M., Assiri, M., & Park, H. J. (2023). Earthquake-induced liquefaction hazard mapping at national-scale in Australia using deep learning techniques. *Geoscience Frontiers*, *14*(1), 101460. https://doi.org/10.1016/j.gsf.2022.101460

Kim, H. S. (2023). Geospatial data-driven assessment of earthquake-induced liquefaction impact mapping using classifier and cluster ensembles. *Applied Soft Computing*, *140*, 110266. https://doi.org/10.1016/j.asoc.2023.110266

Kwak, D. Y., Ahdi, S. K., Wang, P., Zimmaro, P., Brandenberg, S. J., & Stewart J. P. (2021). *Web portal for shear wave velocity and HVSR databases in support of site response research and applications.* UCLA Geotechnical Engineering Group. https://doi.org/10.21222/C27H0V

Lehner, B., & Grill G. (2013). Global river hydrography and network routing: baseline data and new approaches to study the world's large river systems. *Hydrological Processes,* *27*(15): 2171–2186. https://doi.org/10.1002/hyp.9740

Lewis, S. (2009). *Hydrologic Sub-basins of Greenland (NSIDC-0371, Version 1)* [Dataset]. NASA National Snow and Ice Data Center Distributed Active Archive Center. https://doi.org/10.5067/DT9T7DPD7HBI

Lin, A., Wotherspoon, L., Bradley, B., & Motha, J. (2021). Evaluation and modification of geospatial liquefaction models using land damage observational data from the 2010–2011 Canterbury Earthquake Sequence. *Engineering Geology*, *287*, 106099. https://doi.org/10.1016/j.enggeo.2021.106099

Land Information New Zealand (LINZ). (2020). *NZ River Centrelines (Topo, 1:50k)* [Dataset]. LINZ. https://data.linz.govt.nz/layer/50327-nz-river-centrelines-topo-150k/

Maurer, B. W., & Sanger, M. D. (2023). Why "AI" models for predicting soil liquefaction have been ignored, plus some that shouldn't be. *Earthquake Spectra*, *39*(3), 1883-1910. https://doi.org/10.1177/87552930231173711

Maurer, B. W., Green, R. A., & Taylor, O. D. S. (2015a). Moving towards an improved index for assessing liquefaction hazard: Lessons from historical data. *Soils and foundations*, *55*(4), 778-787. https://doi.org/10.1016/j.sandf.2015.06.010

Maurer, B. W., Green, R. A., Cubrinovski, M., & Bradley, B. A. (2015b). Assessment of CPT-based methods for liquefaction evaluation in a liquefaction potential index framework. *Géotechnique*, *65*(5), 328-336. https://doi.org/10.1680/geot.SIP.15.P.007

Maurer, B. W., Green, R. A., van Ballegooy, S., & Wotherspoon, L. (2019). Development of region-specific soil behavior type index correlations for evaluating liquefaction hazard in Christchurch, New Zealand. *Soil Dynamics and Earthquake Engineering*, *117*, 96-105. https://doi.org/10.1016/j.soildyn.2018.04.059

Maurer, B. W., Geyin, M., & van Ballegooy, S. (2025). Cost-benefit analysis of liquefaction mitigation for lightweight residential structures on shallow foundations. In Review. *Journal of Geotechnical and Geoenvironmental Engineering*. In Press. https://doi.org/10.1061/JGGEFK.GTENG-13142

McNeill S.J., Lilburne L.R., Carrick S., Webb T.H., & Cuthill T. (2018). Pedotransfer functions for the soil water characteristics of New Zealand soils using S-map information. *Geoderma,* *326*(15), 96-110. https://doi.org/10.1016/j.geoderma.2018.04.011

Messager, M. L., Lehner, B., Grill, G., Nedeva, I., & Schmitt, O. (2016). Estimating the volume and age of water stored in global lakes using a geo-statistical approach. *Nature Communications,* *7*, 13603. https://doi.org/10.1038/ncomms13603

NASA. (2020) *Distance to nearest coastline.* [Dataset] NASA Ocean Biol Process Group (OBPG) 2. https://oceancolor.gsfc.nasa.gov/resources/docs/distfromcoast/

Nash, J. E., & Sutcliffe, J. V. (1970). River flow forecasting through conceptual models part I—A discussion of principles. *Journal of Hydrology*, *10*(3), 282-290. https://doi.org/10.1016/0022-1694(70)90255-6

National Institute of Building Sciences. (1997). *Earthquake Loss Estimation Methodology: HAZUS Technical Manual* (Vol. 1). Federal Emergency Management Agency.

New Zealand Earthquake Commission (EQC). (2016). *New Zealand Geotechnical Database (NZGD)* [Dataset]. https://www.nzgd.org.nz/

Nobre, A. D., Cuartas, L. A., Hodnett, M., Rennó, C. D., Rodrigues, G., Silveira, A., & Saleska, S. (2011). Height Above the Nearest Drainage–a hydrologically relevant new terrain model. *Journal of Hydrology, 404*(1-2), 13-29. https://doi.org/10.1016/j.jhydrol.2011.03.051

Rasanen, R.A., Geyin, M., & Maurer, B.W. (2023). Select liquefaction case histories from the 2001 Nisqually, Washington earthquake: A digital dataset and assessment of model performance. *Earthquake Spectra*, *39*(3): 1534-1557. https://doi.org/10.1177/87552930231174244

Rasanen, R., Geyin, M., Sanger, M. D., & Maurer, B. W. (2024). *A database of cone penetration tests from the Cascadia Subduction Zone* [Dataset]. DesignSafe-CI. https://doi.org/10.17603/ds2-snvw-jv27

Rateria, G., Geyin, M., & Maurer, B. W. (2024). *CPT-Based Liquefaction Case Histories from Global Earthquakes: A Digital Dataset* [Dataset]. DesignSafe-CI. https://doi.org/10.17603/ds2-8hvd-hd43

Rashidian, V., & Baise, L. G. (2020). Regional efficacy of a global geospatial liquefaction model. *Engineering Geology*, *272*, 105644. https://doi.org/10.1016/j.enggeo.2020.105644

Rashidian, V., & Gillins, D. T. (2018). Modification of the liquefaction potential index to consider the topography in Christchurch, New Zealand. *Engineering Geology*, *232*, 68-81. https://doi.org/10.1016/j.enggeo.2017.11.010

Rathje, E., Dawson, C. Padgett, J.E., Pinelli, J.-P., Stanzione, D., Adair, A., Arduino, P., Brandenberg, S.J., Cockerill, T., Dey, C., Esteva, M., Haan, Jr., F.L., Hanlon, M., Kareem, A., Lowes, L., Mock, S., & Mosqueda, G. (2017). DesignSafe: A new cyberinfrastructure for natural hazards engineering. *Natural Hazards Review, 18*(3). https://doi.org/10.1061/(ASCE)NH.1527-6996.0000246

Regione Emilia-Romagna. (2024). *Banca dati prove geognostiche - Prove penetrometriche numeriche (punti)* [Dataset]. Geoportale. https://geoportale.regione.emilia-romagna.it/catalogo/dati-cartografici/informazioni-geoscientifiche/geologia/banca-dati-geognostica/layer-3

RGI Consortium. (2023). *Randolph Glacier Inventory - A Dataset of Global Glacier Outlines (NSIDC-0770, Version 7).* [Dataset]. National Snow and Ice Data Center. https://doi.org/10.5067/F6JMOVY5NAVZ

Robertson, P. K. (2009). Interpretation of cone penetration tests—a unified approach. *Canadian geotechnical journal*, *46*(11), 1337-1355. https://doi.org/10.1139/T09-065

Sanger, M. D., Geyin, M., Shin, A., & Maurer, B. W. (2024a). *A database of cone penetration tests from North America* [Dataset]. DesignSafe-CI. https://doi.org/10.17603/ds2-gqjm-t836

Sanger, M. D., Geyin, M., Maurer, B. W. (2024b). *Mechanics-informed machine learning for geospatial modeling of soil liquefaction: Global model map products for LPI, LPIish, and LSN* [Dataset]. DesignSafe-CI. https://doi.org/10.17603/ds2-c0z7-hc12

Sanger, M. D., Geyin, M., Maurer, B. W. (2024c). *Mechanics-informed machine learning for geospatial modeling of soil liquefaction: New Zealand model map products for LPI, LPIish, and LSN* [Dataset]. DesignSafe-CI. https://doi.org/10.17603/ds2-hx54-sn38

Sanger, M. D., Geyin, M., Maurer, B. W. (2024d). *Mechanics-informed machine learning for geospatial modeling of soil liquefaction: Example model implementation in a Jupyter Notebook* [Dataset]. DesignSafe-CI. https://doi.org/10.17603/ds2-sp3e-dp21

Shangguan, W., Hengl, T., Mendes de Jesus, J., Yuan, H., & Dai, Y. (2017). Mapping the global depth to bedrock for land surface modeling. *Journal of Advances in Modeling Earth Systems, 9*(1), 65-88. https://doi.org/10.1002/2016MS000686

Smirnov, N. V. (1939). On the estimation of the discrepancy between empirical curves of distribution for two independent samples. *Bull. Math. Univ. Moscou, 2*(2), 3-14.

Steinbach, M. (2000). A comparison of document clustering techniques. In *KDD Workshop on Text Mining, 2000*.

Taftsoglou, M., Valkaniotis, S., Papathanassiou, G., & Karantanellis, E. (2023). Satellite imagery for rapid detection of liquefaction surface manifestations: The case study of Türkiye–Syria 2023 earthquakes. *Remote Sensing, 15*, 4190. https://doi.org/10.3390/rs15174190

Todorovic, L., & Silva V. (2022). A liquefaction occurrence model for regional analysis. *Soil Dynamics and Earthquake Engineering, 161*(1), 107430. https://doi.org/10.1016/j.soildyn.2022.107430

Toprak, S., Nacaroglu, E., van Ballegooy, S., Koc, A. C., Jacka, M., Manav, Y., Torvelainen, E., & O'Rourke, T. D. (2019). Segmented pipeline damage predictions using liquefaction vulnerability parameters. *Soil Dynamics and Earthquake Engineering*, *125*, 105758. https://doi.org/10.1016/j.soildyn.2019.105758

Ulmer, K. J., Zimmaro, P., Brandenberg, S. J., Stewart, J. P., Hudson, K. S., Stuedlein, A. W., Jana, A., Dadashiserej, A., Kramer, S. L., Cetin, K. O., Can, G., Ilgac, M., Franke, K. W., Moss, R. E. S., Bartlett, S. F., Hosseinali, M., Dacayanan, H., Kwak, D. Y., Stamatakos, J., Mukherjee, J., Salman, U., Ybarra, S., & Weaver, T. (2023). *Next-Generation Liquefaction* [Database, Version 2]. Next-Generation Liquefaction Consortium. https://doi.org/10.21222/C23P70

Upadhyaya, S., Green, R. A., Rodriguez-Marek, A., & Maurer, B. W. (2023). True Liquefaction Triggering Curve. *JGGE*, *149*(3), 04023005. https://doi.org/10.1061/JGGEFK.GTENG-11126

USGS. (2019). *Cone Penetration Testing (CPT)* [Dataset]. Earthquake Hazards Program. https://www.usgs.gov/programs/earthquake-hazards/science/cone-penetration-testing-cpt#overview

Van Ballegooy, S., Malan, P., Lacrosse, V., Jacka, M. E., Cubrinovski, M., Bray, J. D., O'Rourke, T. D., Crawford, S.A., & Cowan, H. (2014). Assessment of liquefaction-induced land damage for residential Christchurch. *Earthquake Spectra*, *30*(1), 31-55. https://doi.org/10.1193/031813EQS070M

Wackernagel, H. (2003). *Multivariate geostatistics: an introduction with applications*. Springer Science & Business Media.

Washington State Department of Natural Resources (DNR). (n.d.). *Washington Geologic Information Portal*. Washington State DNR. https://geologyportal.dnr.wa.gov

Westerhoff, R., White, P., & Miguez-Macho, G. (2018). Application of an improved global-scale groundwater model for water table estimation across New Zealand. *Hydrology and Earth System Sciences, 22*(12), 6449-6472. https://doi.org/10.5194/hess-22-6449-2018

Yost, K. M., Green, R. A., Upadhyaya, S., Maurer, B. W., Yerro-Colom, A., Martin, E. R., & Cooper, J. (2021). Assessment of the efficacies of correction procedures for multiple thin layer effects on cone penetration tests. *Soil Dynamics and Earthquake Engineering*, *144*, 106677. https://doi.org/10.1016/j.soildyn.2021.106677

Zhu, J., Baise, L. G., & Thompson, E. M. (2017). An updated geospatial liquefaction model for global application. *Bulletin of the Seismological Society of America*, *107*(3): 1365-1385. https://doi.org/10.1785/0120160198

Zimmaro, P., Nweke, C. C., Hernandez, J. L., Hudson, K. S., Hudson, M. B., Ahdi, S. K., Boggs, M. L., Davis, C. A., Goulet, C. A., Brandenburg, S. J., Hudnut., K. W., & Stewart, J. P. (2020). Liquefaction and related ground failure from July 2019 Ridgecrest earthquake sequence. *Bulletin of the Seismological Society of America*, *110*(4), 1549-1566. https://doi.org/10.1785/0120200025